\documentclass[numberedappendix]{emulateapj}
\usepackage{lscape}
\bibpunct{(}{)}{;}{a}{}{,}

\slugcomment{Accepted to the ApJ}

\shorttitle{ {\it HST} Bump DOG Morphologies}
\shortauthors{Bussmann et al.}

\begin{document}

\title{$HST$ Morphologies of $\lowercase{z} \sim 2$ Dust-Obscured Galaxies II: Bump
Sources}

\author{R. S. Bussmann\altaffilmark{1}, Arjun Dey\altaffilmark{2}, J.
Lotz\altaffilmark{2,3}, L.
Armus\altaffilmark{4}, M. J. I.
Brown\altaffilmark{6}, V.  Desai\altaffilmark{4}, P.
Eisenhardt\altaffilmark{7}, J. Higdon\altaffilmark{8}, S.
Higdon\altaffilmark{8}, B. T.  Jannuzi\altaffilmark{2}, E. Le
Floc'h\altaffilmark{9}, J.  Melbourne\altaffilmark{10}, B. T.
Soifer\altaffilmark{4,10}, D. Weedman\altaffilmark{11}}

\altaffiltext{1}{Steward Observatory, Department of Astronomy, University of
Arizona, 933 N. Cherry Ave., Tucson, AZ 85721; rsbussmann@as.arizona.edu}
\altaffiltext{2}{National Optical Astronomy Observatory, 950 N. Cherry Ave., Tucson, AZ 85719}
\altaffiltext{3}{NOAO Leo Goldberg Fellow}
\altaffiltext{4}{Spitzer Science Center, California Institute of Technology, MS
220-6, Pasadena, CA 91125}
\altaffiltext{6}{School of Physics, Monash University, Clayton, Victoria 3800,
Australia}
\altaffiltext{7}{Jet Propulsion Laboratory, California Institute of Technology,
MC 169-327, 4800 Oak Grove Drive, Pasadena, CA 91109}
\altaffiltext{8}{Georgia Southern University, P.O. Box 8031, Statesboro, GA}
\altaffiltext{9}{Laboratoire AIM-Paris-Saclay, CEA - CNRS -Universite Paris
Diderot, CEA Saclay, Gif-sur-Yvette, 91191, FRANCE}
\altaffiltext{10}{Caltech Optical Observatories, California Institute of
Technology, Pasadena, CA 91125}
\altaffiltext{11}{Astronomy Department, Cornell University, Ithaca, NY 14853}

%\newpage

\begin{abstract}

%The major merger origin for ultra-luminous infrared galaxy (ULIRG) activity in
%the local universe is based in large part on measurements of their rest-frame
%optical morphological properties.  Similar analyses at high redshift have been
%hampered by small sample sizes and systematic uncertainties introduced from
%the use of non-uniform methodology as well as datasets of varying
%sensitivities.  To help rectify this situation, 

We present {\it Hubble Space Telescope} ({\it HST}) imaging of 22 ultra-luminous infrared galaxies (ULIRGs) at $z\approx2$ with extremely red $R-[24]$ colors (called dust-obscured galaxies, or DOGs) which have a local maximum in their spectral energy distribution (SED) at rest-frame 1.6$\mu$m associated with stellar emission.  These sources, which we call ``bump DOGs'', have star-formation rates of $400-4000~M_\sun~$yr$^{-1}$ and have redshifts derived from mid-IR spectra which show strong polycyclic aromatic hydrocarbon emission --- a sign of vigorous on-going star-formation.  Using a uniform morphological analysis, we look for quantifiable differences between bump DOGs, power-law DOGs ({\it Spitzer}-selected ULIRGs with mid-IR SEDs dominated by a power-law and spectral features that are more typical of obscured active galactic nuclei than starbursts), sub-millimeter selected galaxies (SMGs), and other less-reddened ULIRGs from the {\it Spitzer} extragalactic First Look Survey (XFLS).  Bump DOGs are larger than power-law DOGs (median Petrosian radius of $8.4\pm2.7$~kpc vs.  $5.5\pm2.3$~kpc) and exhibit more diffuse and irregular morphologies (median $M_{20}$ of $-1.08\pm0.05$ vs. $-1.48\pm0.05$).  These trends are qualitatively consistent with expectations from simulations of major mergers in which merging systems during the peak star-formation rate period evolve from $M_{20}=-1.0$ to $M_{20}=-1.7$.  Less obscured ULIRGs (i.e., non-DOGs) tend to have more regular, centrally peaked, single-object morphologies rather than diffuse and irregular morphologies.  This distinction in morphologies may imply that less obscured ULIRGs sample the merger near the end of the peak star-formation rate period. Alternatively, it may indicate that the intense star-formation in these less-obscured ULIRGs is not the result of a recent major merger.

% a significant fraction of the high redshift ULIRG population
% is associated with more quiescent modes of formation such as smooth gas inflow
% or the accretion of small satellites.  

%We analyze the morphologies of these objects and compare
%them with a set of 31 ``power-law DOGs'' that have .  
%On the other hand, a significant population of compact, single-object SMG and
%bump sources exists and is difficult to reconcile with the typical
%morphological progression of simulated major mergers.  These results suggest
%that the morphological properties of high redshift ULIRGs are diverse;
%successful models for the formation of such systems must account for this
%apparent morphological complexity.

\end{abstract}


%\newpage

\section{Introduction} \label{sec:intro} 

The discovery of a strong correlation between the stellar bulge mass and the
central super-massive black hole (SMBH) mass of galaxies
\citep[e.g.,][]{1998AJ....115.2285M} has led to detailed theoretical models in
which the growth of SMBHs and their host galaxies occur (nearly) simultaneously
during a brief period of intense, merger-driven activity
\citep[e.g.,][]{2006ApJS..163....1H}.  In these models, the nature of the
connection between SMBHs and their host galaxies has important implications for
the evolution of massive galaxies.  
%Understanding the physical mechanisms that govern the formation and evolution
%of massive galaxies is a key goal of current studies of galaxy evolution.  

The observational foundation of this evolutionary link between SMBHs and their
host galaxies was established by studies of ultra-luminous infrared galaxies (ULIRGs)
% \citep[ULIRGs][]{1996ARA&A..34..749S} 
identified in the local universe using
{\it InfraRed Astronomical Satellite} \citep[see, e.g.,][]{1984ApJ...278L...1N,1996ARA&A..34..749S} data.  ULIRGs
are systems whose spectral energy distributions (SEDs) are dominated by dust
emission at infrared (IR) wavelengths \citep{1986ApJ...303L..41S} and whose
morphologies tend to show evidence for recent or on-going major merger activity
that has been linked to the formation of active galactic nuclei (AGN) and
quasars \citep{1988ApJ...325...74S}.  Although ULIRGs in the local universe are
too rare to contribute significantly to the bolometric luminosity density,
recent studies with the {\it Spitzer Space Telescope} have shown that they
become increasingly important at higher redshifts
\citep[e.g.][]{2005ApJ...632..169L,2009A&A...496...57M}.  To understand the
physical mechanisms that drive massive galaxy evolution, it is essential to
identify and study high-redshift ($z > 1$),  dusty, luminous galaxies that show
signs of concurrent AGN and starburst activity.

%In the cold dark matter paradigm, structure forms in a hierarchical manner:
%the smallest objects are assembled first, and massive objects are built up
%through mergers over time \citep{1978MNRAS.183..341W}.  Although this theory
%of structure formation has many successes, detailed simulations of the
%assembly of baryons into galaxies overpredict the abundance of high-luminosity
%galaxies.

Efforts to identify high-redshift ULIRGs have been increasingly fruitful over the
last two decades.  In particular, blank-field surveys at
sub-millimeter or millimeter wavelengths have identified dusty and rapidly star-forming galaxies, the so-called sub-millimeter
galaxies \citep[SMGs; e.g.][]{1997ApJ...490L...5S,2006MNRAS.372.1621C}.  More
recently, the advent of the Multiband Imaging Photometer for Spitzer
\citep[MIPS;][]{2004ApJS..154...25R} on board the {\it Spitzer Space Telescope}
has allowed for the identification of sources which are bright at mid-IR
wavelengths but faint in the optical
\citep[e.g.][]{2004ApJS..154...60Y,2008ApJ...672...94F,2008ApJ...677..943D,2009ApJ...692..422L}.
Follow-up spectroscopy and clustering measurements of both the
sub-millimeter-selected and the {\it Spitzer}-selected populations has
demonstrated that they have similar number densities, redshift distributions,
and clustering properties that indicate they are undergoing an extremely
luminous, short-lived phase of stellar bulge and nuclear black hole growth and
may be the progenitors of the most luminous ($\sim$4$L^*$) present-day galaxies
\citep{2004ApJ...611..725B,2005ApJ...622..772C,2007ApJ...658..778Y,2006ApJ...641L..17F,2008ApJ...677..943D,2008ApJ...687L..65B}.  

One intriguing difference between the the ULIRG samples selected at different wavelengths (as might be expected
based on the selection criteria) is that the mid-IR selected ULIRGs have hotter
dust than the far-IR selected SMGs
\citep{2006ApJ...650..592K,2008MNRAS.384.1597C,2008ApJ...683..659S,2009MNRAS.394.1685Y,2009ApJ...692..422L,2009ApJ...705..184B,2009A&A...508..117F}.
This distinction may be analogous to the warm-dust/cool-dust dichotomy seen in
local ULIRGs, where it has been suggested that warm ULIRGs represent an
important transition stage between cold ULIRGs and quasars
\citep{1988ApJ...328L..35S}.  Furthermore, the mid-IR-selected population shows a range of spectral energy distributions (SEDs), with the brighter sources showing power-law SEDs in the mid-IR (``power-law'' sources), and the fainter ones exhibiting peaks at rest-frame wavelengths near 1.6$\mu$m (the ``bump'' sources). The bump is generally attributed to starlight and 1.2~mm photometry suggests that the ``bump'' sources are dominated by cooler dust than the power-law sources \citep{2005ApJ...632L..13L,2008ApJ...683..659S,2009MNRAS.394.1685Y,2009ApJ...692..422L,2009ApJ...705..184B,2009A&A...508..117F}.
%There are indications based on
%1.2~mm photometry that the mid-IR-selected population may be divided by
%dust temperature into two sub-classes: those whose mid-IR SEDs contain a peak
%near 1.6$\mu$m (``bump'' sources) have cooler dust than those whose SEDs are
%dominated by a power-law in the mid-IR (``power-law'' sources)
\citep{2005ApJ...632L..13L,2008ApJ...683..659S,2009MNRAS.394.1685Y,2009ApJ...692..422L,2009ApJ...705..184B,2009A&A...508..117F}.
Also, the mid-IR spectra of bump sources show strong polycyclic aromatic
hydrocarbon (PAH) emission features typical of star-forming regions
\citep{2007ApJ...658..778Y,2009ApJ...700.1190D,2009ApJ...700..183H}, while
power-law sources have silicate absorption features or are dominated by
continuum emission consistent with obscured AGN
\citep{2005ApJ...622L.105H,2006ApJ...651..101W,2007ApJ...658..778Y}. These
results suggest that there may be a connection between the power source responsible for the
bolometric luminosity of a system and its globally averaged dust temperature.  

%\subsubsection{PAH Strength and Silicate Absorption Depth}\label{sec:pahsil} 
%
%Many of the DOGs, SMGs, and XFLS ULIRGs studied in this paper have mid-IR
%spectroscopy from {\it Spitzer}/IRS.  Bump DOGs typically show strong PAH
%emission features \citep{2009ApJ...700.1190D}, whereas power-law DOGs usually
%show little or no PAH emission but strong silicate absorption
%\citep{2005ApJ...622L.105H,2006ApJ...651..101W}.  For this reason, the
%morphological distinctions associated with the comparison of strong PAH DOGs to
%weak PAH / strong silicate absorption DOGs mirrors those of the bump ---
%power-law DOG dichotomy.  
%
%Finally, only 7 SMGs in the sample studied here have both high-resolution
%imaging and mid-IR spectroscopy \citep{2009ApJ...699..667M}.  Of these 7, all
%are bump sources, 4 have strong PAH emission, and 3 have weak or no PAH
%emission.  Given the small sample size and the range of results, it is
%difficult to draw any strong conclusions from these sources.

Efforts to understand this connection between mid-IR and far-IR selected
high-$z$ ULIRGs within the context of an evolutionary paradigm have been recently
advanced by numerical simulations of galaxy mergers
\citep[e.g.][]{1996ApJ...464..641M,2009arXiv0910.2234N}.  In these models, when
the merging system approaches final coalescence, the star-formation rate (SFR)
spikes and, because it is enshrouded in cold-dust, the system is observed as an
SMG.  As time proceeds, feedback from the growth of a central super-massive
black hole warms the ambient dust and ultimately quenches star-formation.  It
is during this critical period of galaxy evolution when the system is
observable as a {\it Spitzer}-selected ULIRG.  The models predict observable morphological differences
between the various phases of the merger and, in particular, suggest that mergers occupy a distinct morphological phase space during the ``final coalescence" period when the SFR peaks \citep{2008MNRAS.391.1137L,2009arXiv0912.1590L,2009arXiv0912.1593L}. To test these predictions, and in general to understand the physical processes governing galaxy
evolution, it is essential to study the {\it Spitzer}-selected and SMG
populations in detail.

We have embarked on a detailed study of a large sample of extremely
dust-obscured, high-redshift ULIRGs with the goal of understanding
their evolutionary history. Our sample is selected using {\it Spitzer} and 
ground-based optical imaging of the Bo\"otes field of the NOAO Deep 
Wide-Field Survey \citep[NDWFS\footnote{http://www.noao.edu/noaodeep}; Jannuzi et al., in prep.; Dey et al., in prep.][]{1999ASPC..191..111J} to have extreme optical-to-mid-IR
colors $R-[24]\ge 14$ Vega mag ($\approx F_\nu(24\mu{\rm m})/F_\nu(R)
\ge 1000$) and are called Dust-Obscured Galaxies (DOGs). Spectroscopic
redshifts for a subset of the DOGs have been measured using the
Infrared Spectrometer \citep[IRS;][]{2004ApJS..154...18H} on {\it Spitzer} and optical and
near-IR spectrographs at the W. M. Keck Observatory 
\citep{2005ApJ...622L.105H, 2006ApJ...651..101W, 2009ApJ...700.1190D}. DOGs satisfying
$F_\nu(24\mu{\rm m})\ge 0.3$~mJy have a fairly narrow distribution
in redshift ($z\approx 2.0\pm0.5$) and a space density of $\approx
2.8\times 10^{-5} h_{70}^3 {\rm Mpc}^{-3}$ \citep{2008ApJ...677..943D}.
Although rare, these sources are sufficiently luminous that they
contribute up to one-quarter of the total IR luminosity density at
redshift $z\sim 2$ and constitute a substantial fraction of the
ULIRG population at this redshift.

DOGs are the most dust-reddened ULIRGs at $z\approx 2$; similar to
the broader ULIRG population, DOGs exhibit a wide range in SED
stretching from power-law dominated mid-IR SEDs (i.e., ``power-law
DOGs'') to SEDs which exhibit bumps (i.e, ``bump DOGs'').  In
\citet[][hereafter Paper~I]{2009ApJ...693..750B}, we analyzed the
morphologies of 31 of the brightest 24$\mu$m-selected DOGs (all
with $F_{\rm 24\mu m} > 0.8$ mJy) that have power-law mid-IR SEDs.
All of these objects had spectroscopic redshifts and most exhibit
strong 9.7$\mu$m silicate absorption in their IRS spectra 
\citep{2005ApJ...622L.105H, 2006ApJ...651..101W, 2009ApJ...700.1190D}.
The power-law DOGs are nearly always spatially resolved, with
effective radii of $1-5$~kpc, although a few show obvious signs of
merger activity \citep{2008ApJ...680..232D,2009ApJ...693..750B}. $K$-band
adaptive optics imaging (from Keck) of 15 objects has revealed an intriguing 
dependence of size on SED shape: power-law dominated sources are
more compact than 24$\mu$m-faint bump-dominated sources 
\citep{2008AJ....136.1110M,2009AJ....137.4854M}. This is consistent with 
the idea of the bright, power-law DOGs being more AGN dominated.

The primary goal of this paper is to identify any quantifiable morphological
differences between SMGs, {\it Spitzer}-selected bump ULIRGs and other {\it
Spitzer}-selected power-law ULIRGs. We present and analyze new {\it HST} Wide-Field
Planetary Camera 2 \citep[WFPC2][]{1994ApJ...435L...3T} and Near-IR Camera and Multi-object Spectrometer \citep[NICMOS][]{1998ApJ...492L..95T} observations of 19 bump DOGs and 3 more
power-law DOGs. We also assemble a larger sample of $z\approx 2$ ULIRGs,
drawn from Paper I (power-law DOGs) and the literature, with high
spatial resolution imaging data appropriate for morphological
analyses.  In particular, we include a large sample of SMGs 
\citep[from the stufy of][]{2010MNRAS.405..234S} and expand the sample of 
{\it Spitzer}-selected ULIRGs by including those from the eXtragalactic
First Look Survey \citep[XFLS;][]{2008ApJ...680..232D}. 
Our combined dataset contains 103 high-redshift ULIRGs with available 
and fairly comparable {\it HST} data. We present a uniform morphological analysis of these objects and compare the results to the expectations from models for the formation and evolution of these systems.

In section~\ref{sec:data} we detail the sample selection, observations, and
data reduction.  In section~\ref{sec:methods}, we describe our methodology for
measuring photometry and morphologies, including a visual classification
experiment, non-parametric quantities, and GALFIT modeling.
Section~\ref{sec:results} contains the results of this analysis, including a
comparison of SMG, DOG, and simulated merger morphologies.  In
section~\ref{sec:disc}, we discuss the implications of our results.  We summarize
our conclusions in section~\ref{sec:conclusions}.

Throughout this paper we assume $H_0=$70~km~s$^{-1}$~Mpc$^{-1}$, $\Omega_{\rm
m} = 0.3$, and $\Omega_\lambda = 0.7$.  At $z=2$, this results in a spatial
scale of 8.37~kpc/$\arcsec$.

\section{Data}\label{sec:data}

In this section, we describe the new {\it HST} observations of bump DOGs and
the procedure used to reduce them.  We also detail the archival datasets of
power-law DOGs, SMGs, and XFLS ULIRGs used in subsequent sections of this paper.

\subsection{Bump DOGs}\label{sec:bumpdog}

The 22 DOGs presented in this paper were observed with {\it HST} from 2007
December to 2008 May.  All were observed with WFPC2 through the F814W filter
and with the NICMOS NIC2 camera through the F160W filter.
Table~\ref{tab:observations} summarizes the details of the observations.  All
data were processed using IRAF\footnote{IRAF is distributed by the National
Optical Astronomy Observatory, which is operated by the Association of
Universities for Research in Astronomy, Inc., under cooperative agreement with
the National Science Foundation.  http://iraf.noao.edu/}.  The following
sections provide more details about the sample selection and processing of the
WFPC2 and NICMOS images used in this paper.

\begin{deluxetable*}{lllllllll}
\tabletypesize{\footnotesize} 
\tablecolumns{7}
\tablewidth{6.5in}
\tablecaption{Observations}
\tablehead{
\colhead{Source Name} & 
\colhead{ID\tablenotemark{a}} &
\colhead{RA (J2000)} & 
\colhead{DEC (J2000)} & 
\colhead{$z$\tablenotemark{b}} & 
%\multicolumn{2}{c}{Instrument/Filter} &
\colhead{WFPC2/F814W} & 
%\colhead{$t_{\rm exp}^{\rm opt}$} & 
%\colhead{$t_{\rm exp}^{\rm IR}$} & 
\colhead{NIC2/F160W}
}
\startdata
SST24 J142637.3+333025 &  1 & +14:26:37.397 & +33:30:25.82 & 3.200\tablenotemark{c}                    & 2008-02-11 & 2007-12-31 \\
SST24 J142652.4+345504 & 12 & +14:26:52.555 & +34:55:05.53 & 1.91                    & 2008-03-28 & 2008-01-01 \\
SST24 J142724.9+350823 &  4 & +14:27:25.016 & +35:08:24.20 & 1.71                    & 2008-07-02 & 2008-01-14 \\
SST24 J142832.4+340850 &  8 & +14:28:32.476 & +34:08:51.23 & 1.84                    & 2008-07-03 & 2008-01-15 \\
SST24 J142920.1+333023 & 17 & +14:29:20.164 & +33:30:23.59 & 2.01                    & 2008-02-01 & 2008-05-26\tablenotemark{d} \\
SST24 J142941.0+340915 & 13 & +14:29:41.085 & +34:09:15.61 & 1.91                    & 2008-05-21 & 2008-03-15 \\
SST24 J142951.1+342041 &  5 & +14:29:51.163 & +34:20:41.33 & 1.76                    & 2008-01-28 & 2008-01-14 \\
SST24 J143020.4+330344 & 11 & +14:30:20.537 & +33:03:44.45 & 1.87                    & 2008-03-21 & 2008-04-11 \\
SST24 J143028.5+343221 & 21 & +14:30:28.534 & +34:32:21.62 & 2.178\tablenotemark{e}  & 2008-05-07 & 2008-01-15 \\
SST24 J143137.1+334500 &  7 & +14:31:37.080 & +33:45:01.26 & 1.77                    & 2008-05-20 & 2008-04-12 \\
SST24 J143143.3+324944 &  2 & +14:31:43.400 & +32:49:44.38 & ---                    & 2008-02-10 & 2008-03-15 \\
SST24 J143152.4+350029 &  3 & +14:31:52.463 & +35:00:29.44 & 1.50                    & 2008-01-24 & 2008-05-22 \\
SST24 J143216.8+335231 &  6 & +14:32:16.904 & +33:52:32.18 & 1.76                    & 2008-02-01 & 2008-03-16 \\
SST24 J143321.8+342502 & 18 & +14:33:21.890 & +34:25:02.62 & 2.10                    & 2008-05-21 & 2008-01-15 \\
SST24 J143324.2+334239 & 14 & +14:33:24.269 & +33:42:39.55 & 1.91                    & 2008-02-02 & 2008-01-17 \\
SST24 J143331.9+352027 & 15 & +14:33:31.945 & +35:20:27.28 & 1.91                    & 2007-12-25 & 2008-01-14 \\
SST24 J143349.5+334602 & 10 & +14:33:49.585 & +33:46:02.00 & 1.86                    & 2008-03-18 & 2008-01-07 \\
SST24 J143458.8+333437 & 20 & +14:34:58.953 & +33:34:37.57 & 2.13                    & 2008-07-03 & 2008-05-21 \\
SST24 J143502.9+342657 & 19 & +14:35:02.930 & +34:26:58.88 & 2.10                    & 2008-05-09 & 2008-01-15 \\
SST24 J143503.3+340243 & 16 & +14:35:03.336 & +34:02:44.16 & 1.97                    & 2008-02-29 & 2008-01-07 \\
SST24 J143702.0+344631 & 22 & +14:37:02.018 & +34:46:30.93 & 3.04                    & 2008-03-28 & 2007-12-28 \\
SST24 J143816.6+333700 &  9 & +14:38:16.714 & +33:37:00.94 & 1.84                    & 2008-07-03 & 2008-01-14 \\
\tablenotetext{a}{Panel number in Figure~\ref{fig:cutouts1}}
\tablenotetext{b}{Redshift from {\it Spitzer}/IRS \citep{2009ApJ...700.1190D} unless otherwise noted}
\tablenotetext{c}{Redshift from Keck LRIS (Soifer et al., in prep.)}
\tablenotetext{d}{This observation provided no usable data}
\tablenotetext{e}{Redshift from Keck NIRSPEC \citep{2007ApJ...663..204B}}
\enddata                                                                          
\label{tab:observations}
\end{deluxetable*}

\subsubsection{Sample Selection} \label{sec:sample}

A sample of 2603 DOGs was identified by \citet{2008ApJ...677..943D} from the
9.3 deg$^2$ Bo\"{o}tes Field of the NDWFS.  Keck and {\it Spitzer} spectroscopy
have resulted in redshifts of $\approx~100$ DOGs, approximately 60\% of which
have power-law dominated mid-IR SEDs and 40\% have bump SEDs.  These are
objects which have very high intrinsic to observed UV luminosity ratios, on par
with or beyond the most extreme starbursts studied by {\it Spitzer} in the
local universe \citep{2010ApJ...715..986S}.

In \citet[][hereafter Paper~I]{2009ApJ...693..750B}, we analyzed {\it HST}
imaging (program HST-GO10890) of 31 of the brightest DOGs at 24$\mu m$ (all
have $F_{24\mu {\rm m}} > 0.8 \:$mJy) that have power-law mid-IR SEDs and
spectroscopic redshifts based on the 9.7$\mu$m silicate absorption feature,
most likely due to the presence of warm dust heated by an AGN
\citep{2006ApJ...653..101W,2007ApJ...660..167D,2008ApJ...675..960P,2008ApJ...680..119B}.

In this paper, we analyze {\it HST} imaging (program HST-GO11195) of 22 DOGs
that show a bump in their rest-frame mid-IR SED \citep[selected using Arp~220
as a template; for details see][]{2009ApJ...700.1190D}.  This feature indicates
that the mid-IR light is dominated by stellar emission in these sources.
Furthermore, {\it Spitzer} mid-IR spectroscopy has provided redshifts for 20/22
of these sources via identification of PAH emission features commonly
associated with on-going star-formation \citep{2009ApJ...700.1190D}.
Subsequent deeper mid-IR imaging from the {\it Spitzer} Deep Wide-Field Survey
\citep{2009ApJ...701..428A} has revealed that the two sources lacking PAH
features have power-law mid-IR SEDs.  One additional target has a power-law
mid-IR SED (SST24~J143028.5+343221) and was observed by {\it HST} because the
bump source it replaced could not be observed due to scheduling constraints.

\begin{figure}[!tbp] 
\epsscale{1.00} 
\plotone{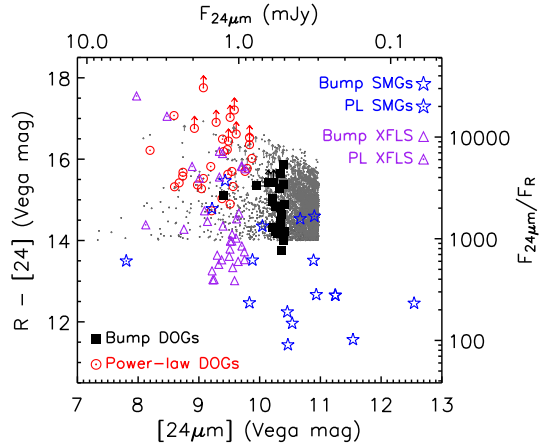}

\caption{$R - [24]$ color vs. 24$\mu$m magnitude distribution for all DOGs in
the NDWFS Bo\"{o}tes field (gray dots).  Arrows indicate $R$-band
non-detections (2$\sigma$ level), and cross symbols highlight power-law
dominated sources.  Also shown are the samples with high-spatial resolution
imaging studied in this paper: power-law DOGs (red circles), bump DOGs (black
squares), SMGs (blue stars), and XFLS ULIRGs (purple triangles).  Power-law
sources tend to be the brightest at 24$\mu$m and the most heavily obscured.
\label{fig:samplecmd}}

\end{figure}

Figure~\ref{fig:samplecmd} shows the $R - [24]$ color and $R$-band magnitude
(Vega system) for the following sources with {\it HST} imaging: bump and
power-law DOGs, SMGs, and XFLS ULIRGs at high redshift.  Following careful
reanalysis of the $R$-band photometry \citep[compared to][with the main
difference being a revised estimate of the sky background
level]{2008ApJ...677..943D}, a few DOGs show $R-[24]$ colors $\approx 0.1~$mag
below the nominal DOG threshold.  We refer to these objects as DOGs in this
paper because they satisfy the essential physical characteristics of DOGs: they
are $z \sim 2$ ULIRGs that are likely to be a highly obscured stage in massive
galaxy evolution.  The bump DOGs in this sample have fainter 24$\mu$m flux
densities and less extreme $R - [24]$ colors than the power-law DOGs.  These
distinctions are qualitatively representative of the photometric properties of
the full sample of 2603 DOGs in the Bo\"otes Field.  

\begin{figure}[!bp]
\epsscale{1.00}
\plotone{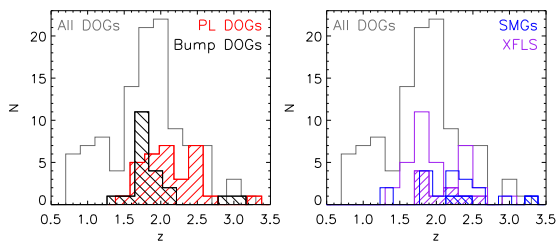}

\caption{ {\it Left}: Redshift distribution of DOGs in the Bo\"{o}tes Field
with spectroscopic redshifts (gray histogram; either from $Spitzer/IRS$ or Keck
DEIMOS/LRIS, Soifer et al. in prep.).  The hatched histograms show the redshift 
distributions of the subset of power-law DOGs (red) and bump DOGs (black) studied 
in this paper. The redshift distribution of bump DOGs is relatively narrow due to selection effects \citep[for details
see][]{2009ApJ...700.1190D}, while power-law DOGs are weighted
towards slightly larger redshifts.  {\it Right}: Redshift distribution of SMGs
(blue histogram) and XFLS ULIRGs (purple histogram) at $z > 1.4$ studied in
this paper.  Hatched regions denote the sub-sample qualifying as power-law
dominated in the mid-IR.  }

\label{fig:zdist}

\end{figure}

Figure~\ref{fig:zdist} shows the redshift distributions of bump DOGs, power-law
DOGs, SMGs, and XFLS ULIRGs with {\it HST} data in comparison to all DOGs in
Bo\"otes with spectroscopic redshifts.  Bump DOGs predominantly lie in a
relatively narrow redshift range of $1.5 < z < 2.1$.  Briefly, this is because
at $z = 1.9$, the strong 7.7$\mu$m PAH feature boosts the 24$\mu$m flux,
pushing sources with weaker continuum into the flux-limited bump DOG
sample \citep[for additional details, see][]{2009ApJ...700.1190D}.

%Shallow X-ray coverage of the Bo\"{o}tes field exists and has yielded a full
%catalog of X-ray sources
%\citep{2005ApJS..161....1M,2005ApJS..161....9K,2006ApJ...641..140B}.  Within a
%2$\arcsec$ search radius, two of the DOGs studied in this paper (SST24
%J143102.2+325152 and SST24 J143644.2+350627) have a single X-ray counterpart,
%and one DOG has two counterparts (SST24 J142644.3+333051).  A full analysis of
%the X-ray data is beyond the scope of this paper, but these basic results
%suggest that most DOGs are either not strong X-ray emitters or are heavily
%obscured.  The latter view is supported both by mid-IR spectral features and
%the fact that this subset of 24$\mu$m bright DOGs shows some of the reddest $R
%- [24]$ colors of the entire DOG population.  Figure~\ref{fig:samplecmd} shows
%the color-magnitude diagram in $R - [24]$ vs.  $[24]$ space for the full DOG
%population in Bo\"{o}tes and highlights the subsample of objects studied in
%this paper.  

\subsubsection{WFPC2 Data} \label{sec:wfpc2} The Wide Field Camera CCD 3 of
WFPC2 was used to image the 22 DOGs in this study.  These observations
consisted of double-orbit data with the F814W filter.  We used a three point
dither pattern (WFPC2-LINE) with a point and line spacing of 0$\farcs$3535 and
a pattern orientation of 45$^\circ$.  Total exposure duration at the nominal
pixel scale of 0$\farcs$1~pix$^{-1}$ was $\approx$3800~sec.  The standard WFPC2
pipeline system was used to bias-subtract, dark-subtract, and flat-field the
images (Mobasher et al., 2002).  MultiDrizzle was then used to correct for
geometric distortions, perform sky-subtraction, image registration, cosmic ray
rejection and final drizzle combination \citep{2002hstc.conf..337K}.  We used a
square interpolation kernel and output pixel scale of
0.075$\arcsec$~pix$^{-1}$, leading to a per-pixel exposure time in the drizzled
image of $\approx$2200~sec.  Typically, a point source with an F814W AB
magnitude of 26.1 may be detected at the 5$\sigma$ level by using a 0\farcs3
diameter aperture.

\subsubsection{NICMOS Data} \label{sec:nicmos} 

Single-orbit data of the DOGs were acquired with the NIC2 camera and the F160W
filter.  We used a two-point dither pattern (NIC-SPIRAL-DITH) with a point
spacing of 0.637$\arcsec$.  The total exposure time per source was $\approx$2700~s.  

We followed the standard data reduction process outlined in the NICMOS data
handbook \citep{nicmos}.  We used the IRAF routine {\tt nicpipe} to pre-process
the data, followed by the {\tt biaseq} task to correct for non-linear bias
drifts and spatial bias jumps.  We then used {\tt nicpipe} a second time to do
flat-fielding and initial cosmic-ray removal.  The IRAF task {\tt pedsky} was
used to fit for the sky level and the quadrant-dependent residual bias.
Significant residual background variation remained after this standard
reduction process.  To minimize these residuals, we followed the procedure
outlined in Paper~I: we constructed an object-masked median sky image based on
all of our NIC2 science frames, scaled it by a spatially constant factor and
subtracted it from each science image.  The scaling factor was computed by
minimizing the residual of the difference between the masked science image and
the scaled sky image.  Mosaicing of the dithered exposures was performed using
{\tt calnicb} in IRAF, resulting in a pixel scale of 0.075$\arcsec$~pix$^{-1}$.
Although the noise varies from image to image, typically a point source with an
F160W AB magnitude of 25.2 may be detected at the 5$\sigma$ level by using a
0\farcs3 diameter aperture.

\subsubsection{Astrometry}\label{sec:astro}

Each WFPC2 and NICMOS image is aligned to the reference frame of the NDWFS,
which itself is tied to the USNO A-2 catalog.  We identify well-detected,
unsaturated sources in the $I$-band NDWFS data overlapping the field of view
(FOV) of each WFPC2/F814W image using Source Extractor
\citep[SExtractor, version 2.5.0,][]{1996A&AS..117..393B}.  The IRAF tasks {\tt wcsctran} and
{\tt imcentroid} are used to convert the RA and DEC values of this list of
comparison sources into WFPC2 pixel coordinates.  Finally, the IRAF task {\tt
ccmap} is used to apply a first order fit which corrects the zero point of the
astrometry and updates the appropriate WCS information in the header of the
WFPC2 image.  The aligned WFPC2 image serves as a reference frame for
correcting the astrometry of the NICMOS image as well as the IRAC images
\citep[since the IRAC images of the Bo\"{o}tes Field are not tied to the USNO
A-2 catalog, but instead to the 2$\mu$m All-Sky Survey frames,
see][]{2009ApJ...701..428A} using a similar procedure.  The properly aligned,
multi-wavelength dataset generally allows for straightforward identification of
the proper counterpart to the MIPS source, since inspection of the four IRAC
channels reveals a single source associated with the 24$\mu$m emission for all
sources.  The absolute uncertainty in the centroid of the IRAC 3.6$\mu$m
emission ranges from 0$\farcs$2-0$\farcs$4.

\subsection{Power-law DOGs}\label{sec:pldogs}

In Paper~I we analyzed {\it HST} imaging
of 31 power-law DOGs at $z > 1.4$.  Although these sources have mid-IR SED
features indicative of obscured AGN, their rest-frame optical morphologies
nearly all show minor ($<30\%$) point-source contributions and significant
emission on scales of 1-5~kpc.  This indicates that the rest-frame optical
light of these sources is produced from stars, rather than AGN.

The NICMOS exposure times and $H$-band luminosities of these sources are
similar to the bump DOGs, facilitating a comparison between the two
populations.  This particular comparison --- between distinct sub-classes of
the most extreme dust-obscured ULIRGs --- is a major aspect of this study.

\subsection{SMG Data}\label{sec:smgdata}

The SMG data used in this paper are {\it HST} NICMOS/F160W imaging of a sample
of 25 SMGs selected from a catalog of 73 SMGs with spectroscopic redshifts
\citep{2005ApJ...622..772C} and were first presented by
\citet{2010MNRAS.405..234S}.  Of the 25 SMGs, 23 have single-orbit NIC2 imaging
from Cycle~12 {\it HST} program GO-9856 \citep{2010MNRAS.405..234S} and an
additional 6 have multi-orbit NIC3 imaging from GOODS-N (Conselice et al. 2010
in prep.).  {\it HST} optical imaging in the F814W filter is also available for
all of these objects.  

In this paper, we focus on the subset of 18 SMGs at $z>1.4$.  Of these 18, all
have NIC2 imaging and three (SMM~J123622.65+621629.7,
SMM~J123632.61+620800.1, and SMM~J123635.59+621424.1) have NIC3 imaging as well.  Although the NIC3
images are significantly deeper, we prefer to use the NIC2 data (each of these
sources is well-detected at S/N$>2$) because of the superior pixel scale of
NIC2 and the unusual shape of the NIC3 PSF.  Some of these sources have optical
{\it HST} imaging with the Advanced Camera for Surveys (ACS), but the S/N
levels are generally insufficient for quantitative analysis and so are not used
in this study.

We obtained the NIC2 images of SMGs from the {\it HST} data archive and reduced
them following the same procedure that is outlined in section~\ref{sec:nicmos}.
Most importantly, the methodology used to analyze the photometry and morphology
of both SMGs and DOGs in this study is identical and is described in
section~\ref{sec:methods}.

\subsection{XFLS Data}\label{sec:xflsdata}

A sample of 33 XFLS ULIRGs at $z > 1.4$ was imaged with {\it HST} NICMOS/F160W
in Cycle~15 as part of program GO10858.  These data and a morphological
analysis of the imaging was presented in \citet{2008ApJ...680..232D}.  We note
that in our study, we use only single-orbit NIC2 data of these objects to
facilitate comparison with the NIC2 images of the other high-$z$ ULIRG
populations studied here, which all have only single-orbit NIC2 data.
Double-orbit imaging is available for nearly 50\% of the sample and in
principle could be used to measure more accurate morphologies of the fainter
objects as well as test for systematic errors in the morphologies resulting
from low S/N.  The data were obtained from the {\it HST} data archive, reduced,
and analyzed using the same methodology that was applied to DOGs and SMGs.

\section{Methodology}\label{sec:methods}

In this section, we describe our methods to measure photometry as well as
visual, non-parametric, and GALFIT morphologies.

\subsection{Photometry}\label{sec:photo}

We perform 2$\arcsec$ diameter aperture photometry on each DOG in both the
NICMOS and WFPC2 images, choosing the center of the aperture to be located at
the peak flux pixel in the NICMOS images.  Foreground and background objects
are identified and removed using SExtractor (see Section~\ref{sec:nonpar}).
The sky level is derived using an annulus with an inner diameter of 2$\arcsec$
and a width of 2$\arcsec$.  In cases where the flux density radial profile did
not flatten at large radii, the appropriate sky value was determined by
trial-and-error.   Photometric uncertainty was computed by measuring the
sigma-clipped root-mean-square of fluxes measured in $N$ $2\arcsec$ diameter
apertures, where $N \approx 10$ and $N \approx 100$ for the NICMOS and WFPC2
images, respectively.  We verified the accuracy of our WFPC2 photometric
zeropoints by comparing well-detected, non-saturated sources common to both the
WFPC2/F814W and NDWFS/$I$-band imaging.  Photometric measurements of the DOGs
are presented in Table~\ref{tab:sources}.

%We compute 4$\arcsec$ diameter aperture photometry in the NDWFS $B_W$, $R$, and
%$I$ images centered on the IRAC 3.6 $\mu$m centroid of emission.  Sky
%background levels were computed in a 3$\arcsec$ wide annulus with an inner
%diameter of 4$\arcsec$.  Limiting magnitudes were determined by measuring the
%flux within a 4$\arcsec$ aperture at several sourceless locations near the DOG
%and computing the rms variation of the flux values.

\subsection{Morphology}\label{sec:morphmeth}

To analyze the morphologies of the bump DOGs, we follow a similar procedure to
that outlined in Paper~I.  Here we summarize the three
different, complementary approaches used in analyzing the morphology of the
DOGs in our sample: a visual classification experiment, multi-component GALFIT
modeling, and non-parametric quantification.  

\subsubsection{Visual Classification} \label{sec:visualmeth}

For this paper, our visual classification experiment differed significantly from
Paper~I.  The goal of the original experiment outlined in Paper~I was to
determine if DOGs could be distinguished from normal field galaxies based on a
visual classification.  This proved difficult to quantify due to the faintness
of DOGs in the rest-frame UV (ACS/WFPC2 images) and the small number of field
galaxies in the rest-frame optical (NICMOS images).

Our new classification experiment is designed specifically to identify
morphological differences found in the NICMOS imaging of bump and power-law
DOGs.  We generated a 5$\arcsec$x5$\arcsec$ cutout image of every DOG with
NICMOS data (both power-law and bump sources, a total of 53 objects) and
arranged them randomly.  Seven of the coauthors classified these objects into
``Regular'', ``Irregular'', or ``Too Faint To Tell''.  In addition to probing
for a difference between bump and power-law DOGs, the mode of the
classifications for each DOG as well as the number of coauthors in agreement
with the mode is useful as a qualititative assessment of the morphology for
comparison with the more quantitative methods discussed below.  
% Seven of the coauthors submitted morphological classifications.  
Results are presented in
Table~\ref{tab:vismorph} and discussed in section~\ref{sec:visual}.  

\subsubsection{Non-parametric Classification} \label{sec:nonparmeth}

A wide variety of tools now exist to quantify the morphologies of galaxies.
Five which frequently appear in the literature are the concentration
index $C$ \citep{1994ApJ...432...75A}, the rotational asymmetry $A$
\citep{1995ApJ...451L...1S}, the residual clumpiness, $S$
\citep{2003ApJS..147....1C}, the Gini coefficient $G$
\citep{2003ApJ...588..218A}, and $M_{20}$ parameter
\citep{2004AJ....128..163L}.  Of these five, $A$ and $S$ have S/N and spatial
resolution requirements that are not satisfied by the existing imaging of the DOGs in
this sample \citep[e.g.,][show that significant type-dependent systematic
offsets in $A$ arise at per-pixel S/N$<5$]{2004AJ....128..163L}.  Therefore,
this analysis is focused on $C$, $G$, and $M_{20}$.

The concentration index $C$ is defined as 
% the ratio of the circular radii containing 20\% and 80\% of the total flux [REPETITIVE - SEE BELOW]
\citep{2000AJ....119.2645B}: 

\begin{equation}
    C = 5 {\rm log10} \left( \frac{ r_{80} }{ r_{20} } \right),
\end{equation}

\noindent where $r_{80}$ and $r_{20}$ are the radii of circular apertures containing
80\% and 20\% of the total flux, respectively.  $G$ was originally introduced
to measure how evenly the wealth in a society is distributed
\citep{glasser1962}.  Recently, \citet{2003ApJ...588..218A} and
\citet{2004AJ....128..163L} applied this method to aid in galaxy
classification: low values imply a galaxy's flux is well-distributed among many
pixels, while high values imply a small fraction of the pixels within a galaxy
account for the majority of the total flux.  $M_{20}$ is the logarithm of the
second-order moment of the brightest 20\% of the galaxy's flux, normalized by
the total second-order moment \citep{2004AJ....128..163L}.  Higher values of
$M_{20}$ indicate multiple bright clumps offset from the second-order moment
center.  Lower values are typical of centrally-dominated systems.

Prior to computing $G$ or $M_{20}$, we first generate a catalog of objects
using SExtractor \citep{1996A&AS..117..393B}.  We use a detection threshold of
3$\sigma$ (corresponding to 23.7~mag~arcsec$^{-2}$) and a minimum detection
area of 15~pixels.  The number of deblending thresholds was 32, and the minimum
contrast parameter for deblending was 0.1.  We found by trial and error that
these parameters minimized the separation of a single galaxy into multiple
components.

For SST24 J143349.5+334602 and SST24 J142652.4+345504, examination of the
F814W-F160W color indicated that a nearby neighbor with similar color should
not be excluded as a foreground/background object.  For both DOGs, we modified
the segmentation map to reflect this.

The final segmentation map produced by SExtractor (and modified in two cases)
is used to mask out foreground/background objects (pixels that are masked out
are simply not used in the remainder of the analysis).  
%The association of pixels with the DOG is determined by the surface brightness
%at the Petrosian radius.  
The center of the image, the ellipticity, and position angle computed by
SExtractor are used as inputs to our morphology code.  This code is written by
J.~Lotz and described in detail in \citet{2004AJ....128..163L}.  Here, we
summarize the relevant information.

Postage stamps of each object in the SExtractor catalog (and the associated
segmentation map) are created with foreground and background objects masked
out.  For each source, we adopt the sky value computed in our photometric
analysis.  Since the isophotal-based segmentation map produced by SExtractor is
subject to the effects of surface brightness dimming at high redshift, pixels
belonging to the galaxy are computed based on the surface brightness at the
elliptical Petrosian radius, $\mu(r_{\rm P})$.  We adopt the usual generalized
definition for $r_{\rm P}$ as the radius at which the ratio of the surface
brightness at $r_{\rm P}$ to the mean surface brightness within $r_{\rm P}$ is
equal to 0.2 \citep{1976ApJ...209L...1P}.  The elliptical $r_{\rm P}$ is
derived from surface brightness measurements within elliptical apertures and
represents the length of the major axis.  Studies have shown that using the
Petrosian radius to select pixels associated with a galaxy provides the most
robust morphological measurements
\citep{2004AJ....128..163L,2008ApJS..179..319L}.  Pixels with surface
brightness above $\mu(r_{\rm P})$ are assigned to the galaxy while those below
it are not.

Using the new segmentation map, we recompute the galaxy's center by minimizing
the total second-order moment of the flux.  A new value of $r_{\rm P}$ is then
computed and a revised segmentation map is used to calculate $G$ and $M_{20}$.
Finally, the morphology code calculates an average S/N per pixel value using
the pixels in the revised segmentation map \citep[Eqs.~1 through 5
in][]{2004AJ....128..163L}.  The S/N per pixel and spatial resolution of each
image is used to estimate the uncertainties in the morphological parameters of
each galaxy.  The uncertainties are derived from the rms variation between
measurements of the same galaxies in GOODS images compared to UDF images
\citep{2006ApJ...636..592L} and assumes that the UDF morphology measurements
are ``truth''.  Results of this analysis are presented in Table~\ref{tab:morph}
and will be discussed in section~\ref{sec:nonpar}.
%We find per-pixel S/N values ranging from $\sim$2-5 for the DOGs.  Reliable
%measurements of $A$ and $S$ require per-pixel S/N$\geq$5
%\citep{2004AJ....128..163L}.  

%One of the most common methods of characterizing galaxy morphologies in the
%literature is to measure the concentration index $C$
%\citep{1994ApJ...432...75A}, the rotational asymmetry $A$
%\citep{1995ApJ...451L...1S}, and the residual clumpiness, $S$
%\citep{2003ApJS..147....1C}.  Given sufficiently high S/N and spatial
%resolution, the $CAS$ system has had demonstrated success in measuring
%morphological parameters and identifying mergers at low
%\citep{2003ApJS..147....1C} and high redshift \citep{2008MNRAS.386..909C}.
%Unfortunately, the objects in our sample do not meet simultaneously the S/N and
%spatial resolution requirements to be reliably placed in $CAS$ space.  Because
%computation of $A$ and $S$ involves differencing two images, the necessary
%per-pixel S/N to measure these parameters reliably is twice as high as those
%that do not involve subtracting images.  In principle, the data are of
%sufficient quality to measure $C$ (see Tab.~\ref{tab:morph}), but in practice
%we find that the inherent assumption of circular symmetry does not apply well
%to the DOGs, making the interpretation of $C$ values difficult.  

\subsubsection{GALFIT Modeling}\label{sec:parmeth}

In Paper~I, we reported the existence of a centrally located, compact component
that was present in the NICMOS images of power-law DOGs but absent in the
ACS/WFPC2 images, signifying the presence of strong central obscuration.  To
quantify this feature, we used GALFIT \citep{2002AJ....124..266P} to model the
2-D light profile of the DOGs.  In this paper, we repeat this procedure on the
bump DOGs with {\it HST} NICMOS data.  Here, we review our methodology.

We choose the size of the fitting region to be 41$\times$41 pixels
(corresponding to angular and physical sizes of 3$\arcsec$ and $\approx$24~kpc,
respectively) because the DOGs are small and have low S/N compared to more
typical applications of GALFIT.  For the same reason, we wish to include only
the minimum necessary components in our model.  We model the observed emission
with three components which are described by a total of 10 free parameters.
The number of degrees of freedom, $N_{\rm DOF}$, is calculated as the
difference of the number of pixels in the image and the number of free
parameters.  Thus, the maximum $N_{\rm DOF}$ is 1671.  Cases where $N_{\rm DOF}
< 1671$ are associated with images where some pixels were masked out because
they were associated with obvious residual instrumental noise.  NIC2 is a
Nyquist-sampled array (0.075$\arcsec$~pix$^{-1}$ compared to 0.16$\arcsec$ FWHM
beam), so the pixels in our image are not completely independent and the
$\chi^2_\nu$ values should be interpreted in a relative sense rather than an
absolute one.

The first element in our GALFIT model is a sky component whose amplitude is
held constant at a value derived from the photometry to yield flat radial
profiles.  The second is an instrumental PSF generated from the TinyTim
software assuming a red power-law spectrum ($F_\nu \propto \nu^{-2}$) as the
object spectrum (Krist and Hook 2004), which can simulate a PSF for NICMOS,
WFPC2, and ACS.  For the NICMOS and WFPC2 images, the DOG is positioned in
nearly the same spot on the camera.  In the case of WFPC2 this is pixel
(132,144) of chip 3 and pixel (155, 164) for NICMOS.  The PSF is computed out
to a size of 3.0$\arcsec$, and for the WFPC2 PSF we oversample by a factor of
1.3 to match the pixel scale of the drizzled WFPC2 images.

The final component is a S\'{e}rsic profile \citep{1968adga.book.....S} where
the surface brightness scales with radius as exp[$-\kappa ( (r/R_{\rm
eff})^{1/n}-1)$], where $\kappa$ is chosen such that half of the flux falls
within $R_{\rm eff}$.  As few constraints as possible were placed so as to
optimize the measurement of the extended flux (i.e., non-point source
component).  In certain cases, the S\'{e}rsic index had to be constrained to be
positive to ensure convergence on a realistic solution.  When fitting the
NICMOS data, the uncertainty image from {\tt calnicb} provides the necessary
information required by GALFIT to perform a true $\chi^2$ minimization.  The
TinyTim NIC2 PSF is convolved with the S\'ersic profile prior to performing the
$\chi^2$ minimization.  The initial guesses of the magnitude, half-light
radius, position angle, and ellipticity were determined from the output values
from SExtractor.  Varying the initial guesses within reasonable values (e.g.,
plus or minus two pixels for the half-light radius) yielded no significant
change in the best-fit model parameters.  The NICMOS centroid was used as the
initial guess for the (x,y) position of both the PSF and extended components.  

A degeneracy potentially exists between our estimates of the point-source
fraction (i.e., relative ratio of PSF component flux to S\'ersic component
flux) and the S\'ersic index.  Fits using models without the PSF
component yield larger reduced $\chi_\nu^2$ values, especially when the point
source fraction in our three-component model was large (see further discussion
in section~\ref{sec:GALFIT}).  In cases where the point source fraction was
small, the no-PSF model had similar parameter values as our fiducial
three-component model, as would be expected.

The results of this GALFIT analysis are presented in Table~\ref{tab:morph} and
will be discussed in section~\ref{sec:GALFIT}.

It is important to note here that NIC2 cannot spatially resolve objects smaller
than 1.3~kpc at $z \approx 2$.  This limit is large enough to encompass a
compact stellar bulge as well as an active galactic nucleus, implying that we
cannot, from these data alone, distinguish between these two possibilities as
to the nature of any central, unresolved component.

\section{Results}\label{sec:results}

In this section, we present our photometry, visual classification,
non-parametric classification, GALFIT modeling, and stellar and dust mass
results.

\subsection{Photometry}\label{sec:photres}

\begin{figure}[!tbp]
\epsscale{1.0}
\plotone{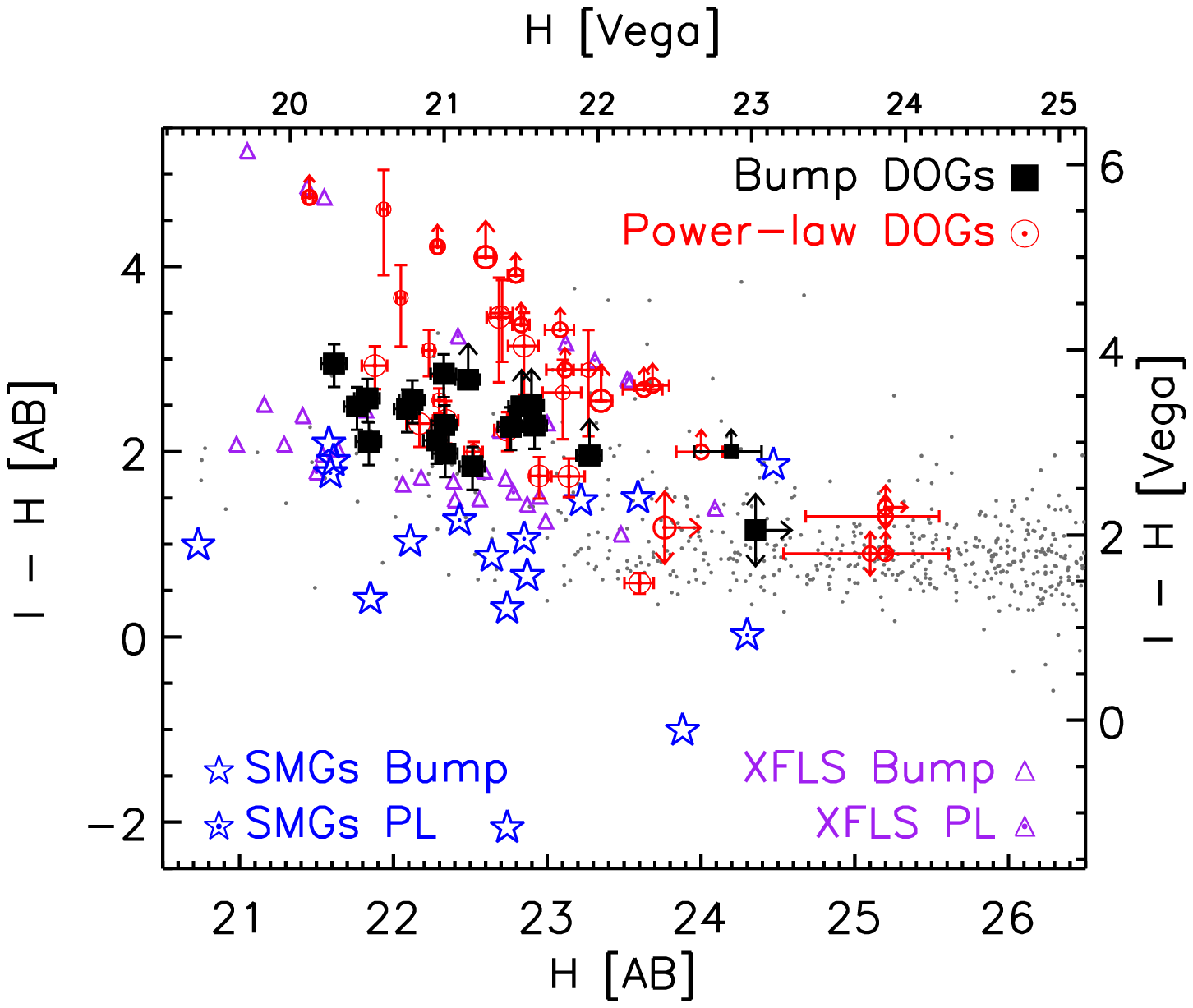}

\caption{ Color-magnitude diagram for bump DOGs, power-law DOGs , SMGs, and
XFLS ULIRGs at $z > 1.4$ (symbols as in Figure~\ref{fig:samplecmd}).  Smaller
symbols indicate objects where the $I$-band measurement has been synthesized
from the $R$-band or $V$-band measurement
\citep{2008ApJ...680..232D,2009ApJ...693..750B}, assuming a power-law of the
form $F_\nu \propto \nu^{-2}$.  Arrows indicate 2-$\sigma$ limits.  Galaxies
spanning the redshift range $1.5 < z < 2.5$ in the HDF-N (Papovich, personal
communication) and HDF-S \citep{2003AJ....125.1107L} are shown with grey dots.
Power-law DOGs have the reddest $I-H$ colors, followed by bump DOGs, XFLS
ULIRGs, and SMGs, which have colors comparable to high-$z$ HDF galaxies. }
\label{fig:imh}

\end{figure}

Table~\ref{tab:sources} presents the photometric information derived from the
{\it HST} imaging.  In Figure~\ref{fig:imh}, we show the $I-H$ vs. $H$
color-magnitude diagram for bump DOGs, power-law DOGs, XFLS ULIRGs, and a
sample of galaxies in the Hubble Deep Field (HDF) whose photometric redshifts
are comparable to DOGs ($1.5 < z_{\rm phot} < 2.5$).  Power-law DOGs tend to be
the reddest sources ($I-H \approx 2-5$ AB mag), followed by bump DOGs ($I-H
\approx 2-3$ AB mag), XFLS ULIRGs ($I-H \approx 1.5-3$), and SMGs, which have
$I-H$ colors similar to high-$z$ galaxies in the HDF ($I-H \approx 0-2$ AB
mag).  SMGs and DOGs (both bump and power-law varieties) are comparably bright
in $H$.  The bluer color of SMGs relative to DOGs at a given $H$-band magnitude
suggests weaker UV flux from DOGs, either due to older stellar populations in
DOGs or a higher dust mass relative to stellar mass in DOGs.

\begin{deluxetable*}{lcccccc}[!tp]
\tabletypesize{\small} 
\tablecolumns{7}
\tablewidth{6.5in}
\tablecaption{Photometric Properties}
\tablehead{
\colhead{} & 
\colhead{$F_{\rm F814W}$} & 
\colhead{$\sigma_{\rm F814W}$} & 
\colhead{$F_{F160W}$} & 
\colhead{$\sigma_{F160W}$} & 
\colhead{$F_{24}$} &
\colhead{$R - [24]$} \\
\colhead{Source Name} & 
\colhead{($\mu$Jy)} & 
\colhead{($\mu$Jy)} & 
\colhead{($\mu$Jy)} &
\colhead{($\mu$Jy)} &
\colhead{(mJy)} &
\colhead{(Vega)}
}
\startdata
SST24 J142637.3+333025 &  0.36  &  0.19  &  0.45  &  0.57  &  0.64  &  $>$15.0 \\
SST24 J142652.4+345504 &  0.24  &  0.15  &  1.78  &  0.36  &  1.29  &     15.0 \\
SST24 J142724.9+350823 &  0.63  &  0.15  &  6.72  &  0.42  &  0.51  &     14.4 \\
SST24 J142832.4+340850 &  0.59  &  0.16  &  ---   &  ---   &  0.52  &     13.9 \\
SST24 J142920.1+333023 &  0.35  &  0.14  &  2.85  &  0.27  &  0.51  &  $>$13.6 \\
SST24 J142941.0+340915 &  0.30  &  0.13  &  2.47  &  0.46  &  0.59  &  $>$14.6 \\
SST24 J142951.1+342041 &  0.55  &  0.16  &  5.30  &  0.52  &  0.60  &  $>$14.9 \\
SST24 J143020.4+330344 &  0.31  &  0.13  &  4.26  &  0.50  &  0.54  &  $>$15.3 \\
SST24 J143028.5+343221 &  0.59  &  0.16  &  4.92  &  0.31  &  1.27  &     14.7 \\
SST24 J143137.1+334500 &  0.18  &  0.14  &  2.67  &  0.37  &  0.57  &     14.3 \\
SST24 J143143.3+324944 &  0.43  &  0.15  &  6.43  &  0.37  &  1.51  &     14.4 \\
SST24 J143152.4+350029 &  0.54  &  0.16  &  8.21  &  0.31  &  0.52  &     14.3 \\
SST24 J143216.8+335231 &  0.51  &  0.15  &  4.24  &  0.37  &  1.28  &  $>$16.1 \\
SST24 J143321.8+342502 &  0.72  &  0.16  &  7.16  &  0.37  &  0.56  &     14.4 \\
SST24 J143324.2+334239 &  0.96  &  0.17  &  6.67  &  0.47  &  0.53  &     13.8 \\
SST24 J143331.9+352027 &  0.66  &  0.17  &  3.58  &  0.32  &  0.60  &     14.3 \\
SST24 J143349.5+334602 &  0.63  &  0.15  &  4.44  &  0.33  &  0.53  &     14.3 \\
SST24 J143458.8+333437 &  0.49  &  0.20  &  5.14  &  0.51  &  0.57  &     14.0 \\
SST24 J143502.9+342657 &  0.24  &  0.13  &  2.52  &  0.63  &  0.50  &     14.1 \\
SST24 J143503.3+340243 &  0.26  &  0.14  &  3.68  &  0.36  &  0.76  &     14.6 \\
SST24 J143702.0+344631 &  0.10  &  0.11  &  0.17  &  0.33  &  0.33  &     14.2 \\
SST24 J143816.6+333700 &  0.68  &  0.16  &  4.22  &  0.22  &  3.28  &     14.8 \\
\enddata
\label{tab:sources}
\end{deluxetable*}

\subsection{Morphologies}

\subsubsection{Visual Classification Results} \label{sec:visual}

From the 7 users who entered classifications of the NICMOS images of DOGs, the
main results can be summarized as follows: power-law DOGs were classified as
irregular (43\%) approximately as frequently as they were classified regular
(42\%), with 15\% being too faint to tell.  In contrast, bump DOGs were
classified as irregular significantly more often than they were classified as
regular (69\% vs. 26\%, with only 5\% being too faint to tell).  These results
can be subdivided into those with very robust classifications (6 or more users
were in agreement), and less robust classifications (fewer than 6 users were in
agreement).  The trends quoted earlier become stronger when considering only
the robust classifications, as the ratio of regular:irregular classifications
for this subset is 1.4:1 and 1:3 for power-law and bump DOGs, respectively.
Table~\ref{tab:vismorph} shows the breakdown of visual classifications with
this additional subdivision.  In Table~\ref{tab:morph} we provide, for each DOG
in this sample, the mode of the classifications as well as how many users were
in agreement with the mode.  Overall, the qualitative morphological assessment
indicates that bump DOGs have irregular, diffuse morphologies more frequently
than power-law DOGs.

\subsubsection{Non-parametric Classification Results} \label{sec:nonpar}

The characterization of galaxy morphologies requires high S/N imaging in order
to provide reliable results.  For non-parametric forms of analysis, typical
requirements are ${\rm S/N_{pixel}} > 2$ and $r_{p}({\rm Elliptical}) > 2
\times {\rm FWHM}$ \citep{2004AJ....128..163L} (hereafter, $r_{\rm P}$
indicates the elliptical petrosian radius).  In the case of the imaging
presented here, ${\rm FWHM} = 0\farcs16$.  None of the 20 bump DOGs in this
study observed with WFPC2 have the per-pixel S/N necessary to compute $r_{\rm
P}$, $G$, $M_{20}$, and $C$.  On the other hand, 18 out of 20 sources have
sufficient S/N in the NICMOS imaging.  Table~\ref{tab:morph} presents the
visual and non-parametric measures of DOG morphologies, including per-pixel
S/N, $r_{\rm P}$, $G$, $M_{20}$, and $C$ values for the NICMOS images.  This
table also includes an estimate of whether the DOG is dominated by a bump or by
a power-law in the mid-IR using IRAC data from \citet{2009ApJ...701..428A} and
the same statistical definition originally used by \citet{2008ApJ...677..943D}.

%\subsubsection{DOG Size Scales}\label{sec:dogsize} 

\begin{deluxetable*}{lcccccc}[!bp]
%\tabletypesize{\small} 
\tablecolumns{7}
\tablewidth{0in}
\tablecaption{Visual Morphological Classifications\label{tab:vismorph}}
\tablehead{
\colhead{} & \multicolumn{2}{c}{Regular}  & \multicolumn{2}{c}{Irregular} &
\multicolumn{2}{c}{Too Faint Too Tell} \\
\colhead{} & \colhead{Agree}  & \colhead{Disagree} & \colhead{Agree} &
\colhead{Disagree} & \colhead{Agree} & \colhead{Disagree}
}
\startdata
Power-law DOGs & 34\% & 9\% & 24\% & 18\% & 6\% & 9\% \\
Bump DOGs      & 16\% & 10\% & 48\% & 21\% & 5\% & 0\% \\
\enddata

\end{deluxetable*}

\begin{figure*}[!tbp]
\epsscale{1.00}
\plotone{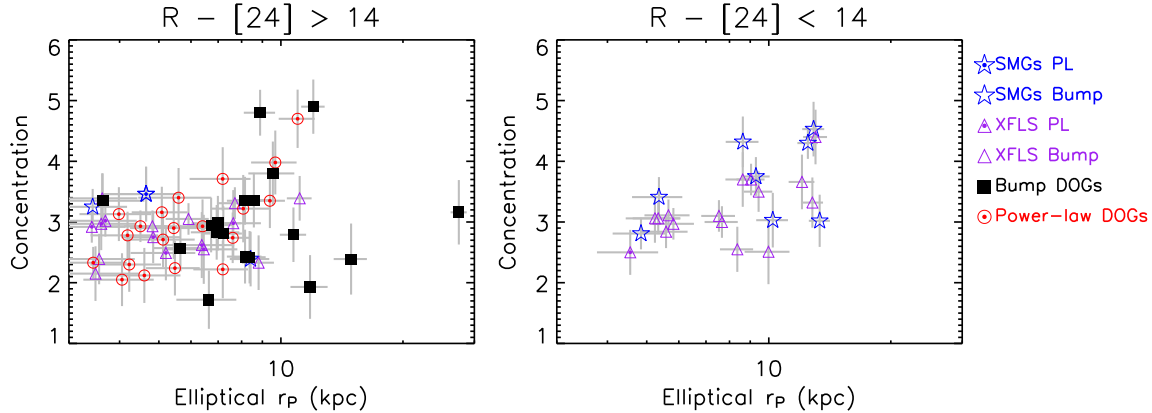}

\caption{ $C$ as a function of $r_{\rm P}$ for $z > 1.4$ ULIRGs (symbols are
the same as in Figure~\ref{fig:samplecmd}).  {\it Left}: Power-law DOGs, bump
DOGs, SMGs that qualify as DOGs, and XFLS ULIRGs at $z>1.4$ that qualify as
DOGs.  Error bars illustrate the typical uncertainty level
given the S/N and spatial resolution associated with the image of each galaxy
\citep{2006ApJ...636..592L}.  Bump DOGs have larger sizes than power-law DOGs.
{\it Right}: Same as left panel, but only for $z > 1.4$ ULIRGs (SMGs and XFLS)
that are not DOGs.  Regardless of sample selection criteria, power-law $z >
1.4$ ULIRGs are significantly smaller than their bump counterparts (median
$r_{\rm P}$ of 5.6~kpc vs.  8.0~kpc, for the total respective populations).}

\label{fig:rpetmag}

\end{figure*}

Figure~\ref{fig:rpetmag} displays $C$ as a function of $r_{\rm P}$ for
power-law DOGs, bump DOGs, SMGs, and XFLS sources.  The error bars indicate the
typical uncertainties in $C$ and $r_{\rm P}$ given the S/N and spatial
resolution associated with the imaging of each galaxy.  The left panel of
Figure~\ref{fig:rpetmag}, focusing only on bump and power-law sources that
qualify as DOGs, shows that bump DOGs have larger sizes (median $r_{\rm P} =
8.4$~kpc, $\sigma_{r_{\rm P}} = 2.7$~kpc) than their power-law counterparts
(median $r_{\rm P} = 5.5$~kpc, $\sigma_{r_{\rm P}} = 2.3$~kpc).  A two-sided
Kolmogorov-Smirnov (KS) indicates only a 1\% chance the two $r_{\rm P}$
distributions are drawn from the same parent distribution.  The right panel of
Figure~\ref{fig:rpetmag} shows SMGs and XFLS sources which are not DOGs.  In
this diagram, almost all sources are bumps, and almost all sources have large
sizes (median $r_{\rm P} = 8.5~$kpc, $\sigma_{r_{\rm P}} = 2.9$~kpc).

When no consideration is given to their $R-[24]$ color, SMGs and XFLS sources
show a similar distinction in their sizes when dividing the samples into bump
(SMG median $r_{\rm P} = 8.6$~kpc, $\sigma_{r_{\rm P}} = 3.3$~kpc; XFLS median
$r_{\rm P} = 7.6$~kpc, $\sigma_{r_{\rm P}} = 2.9$~kpc) and power-law (SMG
median $r_{\rm P} = 4.6$~kpc, $\sigma_{r_{\rm P}} = 4.5$~kpc; XFLS median
$r_{\rm P} = 4.8$~kpc, $\sigma_{r_{\rm P}} = 1.5$~kpc) varieties.  Indeed,
considering all $z>1.4$ ULIRGs regardless of whether they are selected at
mid-IR or sub-mm wavelengths, bump sources (median $r_{\rm P} = 8.4$~kpc,
$\sigma_{r_{\rm P}} = 2.9$~kpc) are significantly larger than their power-law
counterparts (median $r_{\rm P} = 5.6$~kpc, $\sigma_{r_{\rm P}} = 1.9$~kpc),
and a two-sided KS test indicates there is only a 1.3\% chance the two
populations could be drawn randomly from the same parent sample.  This finding
is consistent with results from Keck $K$-band adaptive optics imaging of DOGs
which shows that power-law DOGs are smaller and more concentrated than bump
DOGs \citep{2009AJ....137.4854M}.  One caveat with this result is that the bump
DOG sample is brighter in $H$-band than the power-law DOG sample.  Considering
only the DOGs satisfying $H < 22.5$, the bump and power-law DOGs have similar
sizes ($r_{\rm P} \approx 8$~kpc).  At the faint end ($H > 22.5$), power-law
DOGs are smaller than bump DOGs (5~kpc vs. 8~kpc, respectively).

%\subsubsection{$G$ and $M_{20}$}\label{sec:gm20}  

\begin{figure*}[!tbp]
\epsscale{1.00}
\plotone{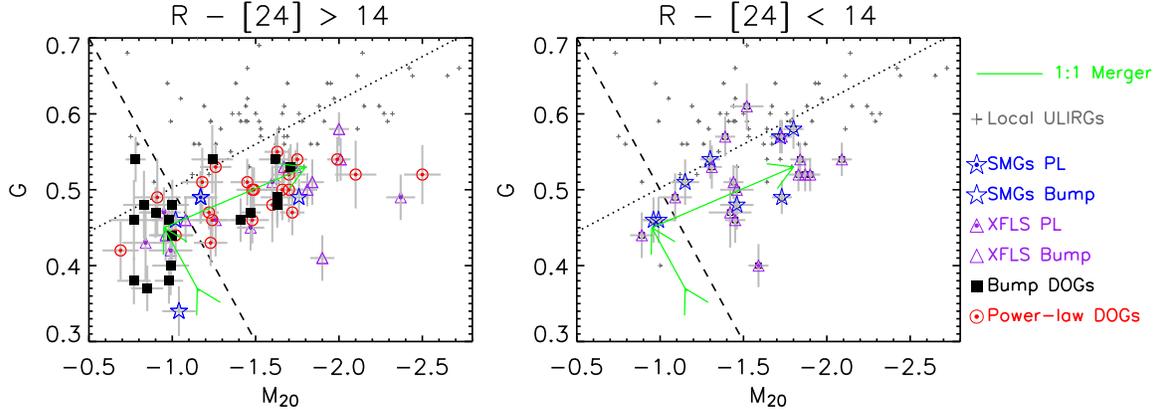}

\caption{Gini coefficient vs. $M_{20}$ derived from NIC2/F160W images of
high-redshift ULIRGs (symbols same as in Figure~\ref{fig:rpetmag}) and local
ULIRGs \citep[gray plus signs,][]{2004AJ....128..163L}.  The evolution of a
typical gas-rich ($f_{\rm gas} = 0.5$) major merger during its peak SFR period
is illustrated by a green vector \citep{2008MNRAS.391.1137L}.  The dashed line
is drawn qualitatively to separate ``diffuse'' and ``single-object''
morphologies and bisects the green vector.  The dotted line shows the
empirically determined (based on measurements of local ULIRGs) demarcation line
above which objects are obvious major mergers \citep{2004AJ....128..163L}.
{\it Left}: Bump DOGs, power-law DOGs, and SMGs and XFLS ULIRGs qualifying as
DOGs.  Within this highly obscured subset of the high redshift ULIRG
population, bump sources are ``diffuse'' (low $G$, high $M_{20}$) more often
than power-law DOGs.  In simulations of major mergers, such morphologies occur
during the early half of the peak SFR period of the merger.  {\it Right}: Same
as left panel, but for SMGs and XFLS $z>1.4$ ULIRGs that are not DOGs.  The
distribution of morphologies for non-DOGs is skewed towards the
``single-object'' region of this diagram.  These objects may occur during the
late stage of the peak SFR period of a major merger, or they may be associated
with more secular evolutionary processes.}

\label{fig:gm20}

\end{figure*}

The distribution in $G-M_{20}$ space derived from NICMOS imaging of power-law
DOGs, bump DOGs, XFLS sources, and SMGs is shown in Figure~\ref{fig:gm20}.  The
error bars indicate the typical uncertainties in $G$ and
$M_{20}$ given the S/N and spatial resolution of the imaging of each galaxy.  A sample of 73 local
ULIRGs ($z < 0.2$) is also shown in this diagram \citep{2004AJ....128..163L},
using data from {\it HST} WFPC2/F814W imaging \citep{2000ApJ...529L..77B}.  The
dotted line separates major mergers from other types of galaxies and is based
on measurements at roughly the same rest-frame wavelength ($\approx
5000-5500$~\AA) of these 73 local ULIRGs \citep{2004AJ....128..163L}.

The left panel of Figure~\ref{fig:gm20} (including all sources that qualify as
DOGs) shows that bump DOGs appear offset to lower $G$ and higher $M_{20}$
values than power-law DOGs.  The median \{$G$, $M_{20}$\} values for bump and
power-law DOGs are \{0.47, -1.08\} and \{0.49, -1.48\}, respectively.  A
two-sided KS test indicates that there is only a 0.5\% chance that the two
$M_{20}$ distributions could have been drawn randomly from the same parent
distribution (the two $G$ distributions have a 10\% chance of being drawn from
the same parent distribution).  These types of morphologies are consistent with
what is seen in simulations of major mergers during the beginning and end
stages, respectively, of the ``final coalescence'' of the merger when the SFR
peaks and begins to turn over \citep{2008MNRAS.391.1137L}.  In the right panel
of Figure~\ref{fig:gm20}, SMGs and XFLS $z>1.4$ ULIRGs that are not DOGs are
shown.  Although nearly all of these sources have bump SEDs, their morphologies
bear a greater resemblence to power-law DOGs than bump DOGs.  The median \{$G$,
$M_{20}$\} values for the non-DOGs are \{0.52, -1.46\}.  

The preceding analysis is largely qualitative in nature.  A more quantitative
approach involves the use of contingency tables, which offer a means to
quantify broad-brush distinctions in the properties of two populations of
objects.   Three properties are tested here: mid-IR SED shape (bump OR
power-law), extent of obscuration ($R-[24] > 14$ OR $R-[24] < 14$), and
morphology (low $G$, high $M_{20}$ OR high $G$, low $M_{20}$).  The division
based on morphology is derived from simulations of major mergers, which
indicate that the high SFR period of a merger is bisected by a line described
by the equation $G = 0.4 M_{20} + 0.9$ \citep{2008MNRAS.391.1137L}.
Table~\ref{tab:contingency} shows the two 2$\times$2 contingency tables that
are needed to account for the three variables used in this analysis.

The first result from this analysis is the paucity of power-law sources in the
non-DOG subset.  There are 29 bump DOGs, 23 bump non-DOGs, 31 power-law DOGs,
and only 1 power-law non-DOGs.  The $2\times2$ contingency table for this
dataset indicates a negligible probability (Fisher Exact p-value $<0.0001$)
that all four sub-populations are drawn randomly from the same parent sample.
Could this be due to a selection effect?  The non-DOG sample comprises ULIRGs
from the XFLS and SMGs.  XFLS sources are selected to have high $F_{\rm 24\mu
m} / F_{\rm 8\mu m}$ flux density ratios, which tends to favor the selection of
bump SEDs over power-law ones.  On the other hand, the XFLS sources are
selected to be very bright at 24$\mu$m ($F_{\rm 24\mu m} > 0.8$~mJy.  At these
24$\mu$m flux densities, power-law sources are more common than bump sources
\citep[e.g.][]{2008ApJ...677..943D}.  SMGs are selected at sub-mm wavlengths,
without any knowledge of the mid-IR SED shape.  Presently, it is not obvious
that either the XFLS ULIRGs or SMGs are affected by the kind of severe
selection effect necessary to produce the observed trends.

The second result from the contingency table data is that, considering only
bump sources, non-DOGs have a much more skewed distribution of morphologies
than DOGs.  Diffuse type morphologies (low $G$, high $M_{20}$) are rare in the
non-DOG population, while in DOGs they occur much more frequently.  A
$2\times2$ contingency table here suggests a very low probability (Fisher Exact
p-value $=0.007$) that blue ($R-[24]<14$) and red ($R-[24]>14$) ULIRGs have
morphologies drawn from the same parent distribution.  Low $G$ and high
$M_{20}$ values suggest irregular and lumpy (less centrally concentrated)
morphologies that could be caused by a clumpy distribution of stars or
significant dust obscuration \citep{2008MNRAS.391.1137L}.  Further discussion
of the implications of this result are deferred to section~\ref{sec:disc}.  

Finally, with the highly obscured subset of ULIRGs (DOGs), there is evidence
that bump DOGs have diffuse type morphologies more commonly than power-law
DOGs.  A $2\times2$ contingency table indicates an extremely low probability
(Fisher Exact p-value $=0.003$) that bump and power-law DOGs have morphologies
drawn from the same parent distribution.  As mentioned earlier, this
distinction is consistent with expectations from simulations of major mergers
during the peak SFR phase of the merger \citep{2008MNRAS.391.1137L}.

\begin{deluxetable}{lcccc}
%\tabletypesize{\small} 
\tablecolumns{5}
\tablewidth{3.5in}
\tablecaption{NICMOS Morphology Contingency Table Data}
\tablehead{
\colhead{} & \multicolumn{2}{c}{$R-[24]<14$}  & \multicolumn{2}{c}{$R-[24]>14$} \\
\colhead{} & 
\colhead{Diffuse\tablenotemark{a}}  &
\colhead{Single-source\tablenotemark{b}} & 
\colhead{Diffuse\tablenotemark{a}} & 
\colhead{Single-source\tablenotemark{b}}
}
\startdata
Power-law &	0 &  1 &  7 & 24 \\
Bump &		3 & 20 & 15 & 14 \\
\enddata
\tablenotetext{a}{$G < 0.4 M_{20} + 0.9$}
\tablenotetext{b}{$G > 0.4 M_{20} + 0.9$}
\label{tab:contingency}
\end{deluxetable}

\subsubsection{GALFIT Results}\label{sec:GALFIT}

The results of our GALFIT analysis of the NICMOS images of the Cycle~16 DOGs
are shown in Table~\ref{tab:morph}, along with 1-$\sigma$ uncertainties in the
best-fit parameters.  Included in this table are point source fractions (ratio
of flux in the point-source component to the total flux of the source),
effective radius of the S\'ersic component ($R_{\rm eff}$), S\'ersic index
($n$), semi-minor to semi-major axis ratio of the S\'ersic component (Axial
Ratio), number of degrees of freedom ($N_{\rm DOF}$), and reduced chi-squared
($\chi_\nu^2$).

\begin{figure*}[!tbp]
\epsscale{1.00}
\plotone{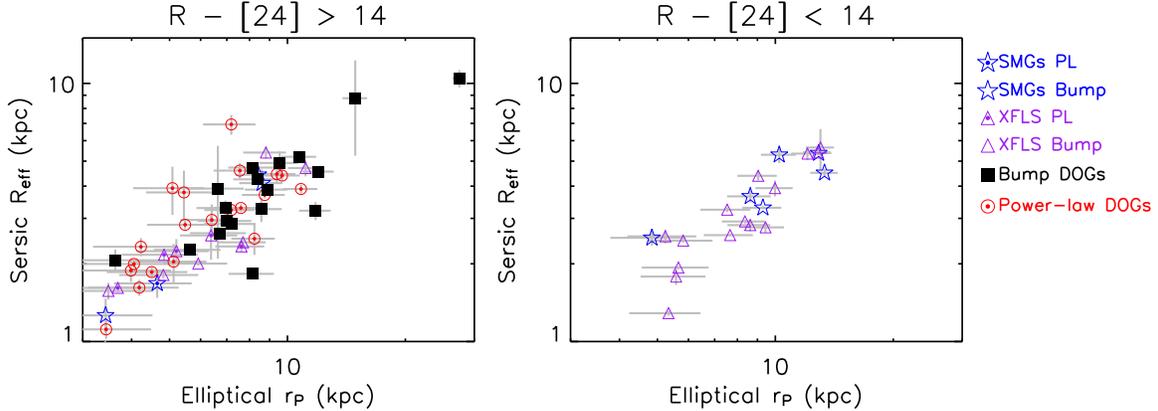}

\caption{Comparison of sizes of $z > 1.4$ ULIRGs (symbols same as in
Figure~\ref{fig:rpetmag}) as determined by the effective radius of the
S\'ersic component from GALFIT modeling ($R_{\rm eff}$) and the elliptical
Petrosian radius ($r_{\rm P}$).  Error bars represent 1$\sigma$ uncertainty
values from GALFIT.  {\it Left}:  Bump DOGs, power-law DOGs, and
SMGs and XFLS ULIRGs qualifying as DOGs.  {\it Right}: SMGs and XFLS ULIRGs
that are not DOGs.  Both size measurements suggest that power-law sources are
on average smaller than bump sources, although a significant population of
compact bump sources exists. }

\label{fig:reffrpet}

\end{figure*}

Figure~\ref{fig:reffrpet} shows a comparison of $R_{\rm eff}$ (the radius
within which half the light is enclosed) and $r_{\rm P}$ (the radius at which
the ratio of the surface brightness at $r_{\rm P}$ to the mean surface
brightness within $r_{\rm P}$ is equal to 0.2) for DOGs, SMGs, and XFLS ULIRGs
at $z > 1.4$.  For bump DOGs and power-law DOGs, the median $R_{\rm eff}$
values are 3.3~kpc and 2.5~kpc, respectively.  Bump sources that are not DOGs
(from the SMG and XFLS samples) have a median effective radius of 3.2~kpc.  One
of the bump DOGs (SST24~J143137.1+334500) has the appearance of an edge-on disk
with a semi-major axis of $3\farcs25$, or 27.5~kpc at its redshift of 1.77.
This extremely large $R_{\rm eff}$ value may imply that this object is in fact
a merger viewed edge-on.  Spatially resolved dynamical information would be
particularly useful for answering this question.

Our measurements of SMG sizes (median $R_{\rm reff}$ value for the full SMG
population of 3.6~kpc) are in broad agreement, given the different methods
used, with those of \citet{2010MNRAS.405..234S}, who find typical half-light
radii of $2.8\pm0.4$~kpc.  For XFLS ULIRGs, \citet{2008ApJ...680..232D} use
GALFIT to find typical effective radii of $2.43\pm0.80$~kpc, consistent with
our results (median $R_{\rm eff}$ of 2.5~kpc).  As an additional consistency
check, a strong correlation is evident between  $R_{\rm eff}$ and $r_{\rm P}$
for all populations.  Note that $r_{\rm P} > R_{\rm eff}$; this is because the
S\'ersic profile is defined such that half of the galaxy's flux is enclosed
within a radius of $r=R_{\rm eff}$, while $r_{\rm P}$ defines the radius at
which the surface brightness is one-fifth the average surface brightness within
$r_{\rm P}$.

Figure~\ref{fig:fpsfn} shows the point source fraction and S\'ersic index for
DOGs, SMGs, and XFLS ULIRGs at $z > 1.4$.  The majority of sources have low
point source fractions (point source fraction $<0.3$) and disk-type
morphologies ($n < 2$).  Studies have found that when a point source
contributes less than 20\% of the total light, it has an insignificant effect
on the measured morphologies  \citep{2010MNRAS.tmp..455P}.  Considering only
DOGs with sufficient S/N to be placed on this diagram (left panel of
Figure~\ref{fig:fpsfn}), 6/28 power-law DOGs and 0/17 bump DOGs have either $n
> 3$ or point source fraction $>0.4$.  Such sources have compact, centrally
dominated morphologies \citep[$n=1$ corresponds to an exponential profile, and
$n=4$ corresponds to a de Vaucouleurs profile;][]{2002AJ....124..266P}.  This
distinction is consistent with the $G$ and $M_{20}$ results in
section~\ref{sec:nonpar}.  

\begin{figure*}[!tbp]
\epsscale{1.00}
\plotone{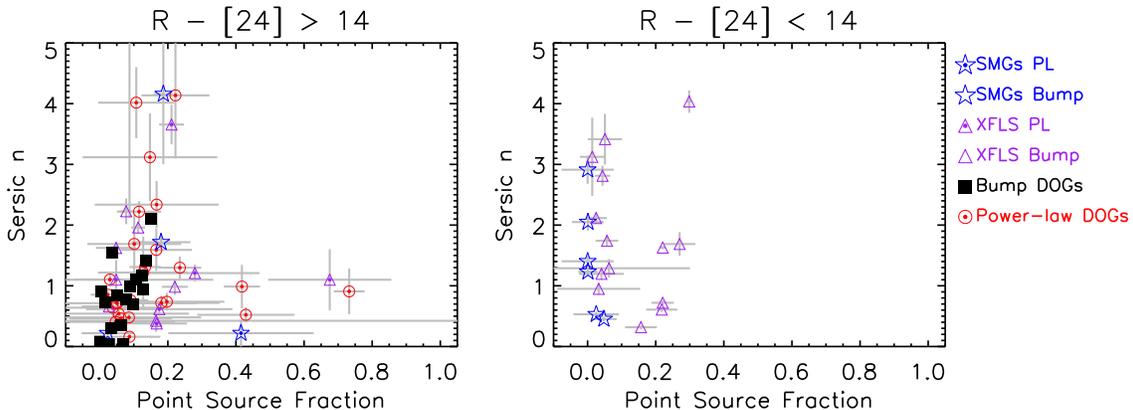}

\caption{S\'ersic index $n$ as a function of point source fraction from GALFIT
modeling (symbols same as in Figure~\ref{fig:reffrpet}).  {\it Left}:
Power-law DOGs, bump DOGs, and SMGs and XFLS ULIRGs qualifying as DOGs.  Aside
from a handful of power-law DOGs with point source fraction $>0.4$ or $n >
2.5$, there is strong overlap between the bump and power-law DOG populations in
this diagram.  {\it Right}:  SMGs and XFLS ULIRGs that do not qualify as DOGs.
In contrast to the DOG populations, there are a number of $n>2.5$ bump sources
from the SMG and XFLS samples.  As in the analysis of the $G$ and $M_{20}$
values, these could represent objects at the end of the peak SFR period, or
they might not be associated with major merger activity at all. }

\label{fig:fpsfn}

\end{figure*}

On the other hand, the distinction between bump and power-law sources is not as
obvious when considering the SMGs and XFLS sources.  For SMGs, 2/3 power-law
and 2/11 bump sources satisfy the compact criteria outlined above, while for
XFLS ULIRGs the respective numbers are 2/6 (power-law sources) and 3/18 (bump
sources).  Further discussion of the distinction between the morphological
properties of bump and power-law DOGs is deferred to section~\ref{sec:disc}.

\section{Discussion: Implications for Models of Massive Galaxy
Evolution}\label{sec:disc}

ULIRG activity in the local universe has been known for some time to result
from a major merger of two gas-rich disk galaxies
\citep[e.g.][]{1987AJ.....94..831A, 1988ApJ...325...74S}.  Material is funneled
towards the center of the system and drives an intense starburst, producing
large amounts of cold dust, and begins to feed a nascent central black hole.
As the merger evolves, ambient gas and dust particles are heated by feedback
processes.  This warm-dust ULIRG stage has been suggested to represent a
transition stage between cold ULIRGs and optically luminous quasars
\citep{1988ApJ...328L..35S}.

Recently, efforts have been made to extend this paradigm to the ultra-luminous
galaxy populations at high-redshift.  One possible hypothesis within this
scenario is that SMGs represent the cold-dust ULIRGs created during the early
stage of the merger, whereas {\it Spitzer}-selected sources represent the
warm-dust ULIRGs formed during the later stages of the merger
\citep[e.g.,][]{2008ApJ...677..943D,2009ASPC..408..411D,2009arXiv0910.2234N}.
This basic picture (that SMGs and {\it Spitzer}-selected ULIRGs are related) is
strengthened by the similarity in the measured clustering strengths of $z
\approx 2$ SMGs, DOGs, and QSOs, which suggest that these populations all
reside in similar mass halos at similar epochs
\citep[e.g.,][]{2008ApJ...687L..65B, 2009ASPC..408..411D}. 

In this section, we test the viability of this scenario using the morphological evidence presented in
section~\ref{sec:results}.  On one hand,
when considering only the most extremely obscured objects (i.e., DOGs), a clear
trend in morphologies emerges.  Bump DOGs are larger (i.e., more spatially
extended) than power-law DOGs ($r_{\rm P} \approx 8$~kpc vs. 5~kpc), more
diffuse ($\{G,M_{20}\} \approx \{0.47, -1.08\}$ vs.  $\{G,M_{20}\} \approx
\{0.49, -1.48\}$), and more irregular (67\% vs. 50\% visually classified as
irregular).  This trend is consistent with expectations from simulations of
major mergers, which indicate that merger morphologies generally evolve from
extended, diffuse, and irregular at the beginning of the peak SFR phase to
compact and regular when star-formation shuts down and the AGN begins to
dominate \citep{2008MNRAS.391.1137L,2009arXiv0910.2234N}.

On the other hand, the less obscured sources (non-DOGs from the SMG and XFLS
sample) show two strong distinctions from their more extreme counterparts.
First, there are very few power-law non-DOGs.  If power-law SEDs are more
frequently associated with objects that are more dust reddened, this may imply
a connection between the amount of extinction of the optical light and the
nature of the power source producing the mid-IR emission.

Second, within the bump population of non-DOGs, there are very few diffuse type
morphologies (low $G$, high $M_{20}$).  The prevalence of bump sources with
``single-object'' morphologies is difficult to understand within the context of
a major merger scenario in which bump sources evolve into power-law sources.
If the bump phase always precedes the power-law phase, there should be very few
bump sources with compact, single-object morphologies.  A number of potential
explanations exist.

Perhaps the most exciting explanation is that high redshift ULIRGs are related
to one another within a single evolutionary scheme driven by major mergers, but
with an additional wrinkle related to the degree of obscuration.  During the
highly dust-obscured period of the merger (represented jointly by both bump and
power-law DOGs), the bump phase typically occurs before the power-law phase.
In contrast, the less obscured sources (SMGs and XFLS ULIRGs) sample the merger
over a broader timescale and so the relationship between bump and power-law
sources is not as obvious.  For example, there may be a significant population
of blue ULIRGs (non-DOGs) that correspond to the systems near the very end of the high SFR period
of the merger when the obscuring column of dust has decreased and UV light can
escape the galaxy.

An alternative, but potentially equally exciting, way to reconcile the
morphological evidence is by appealing to more quiescent modes of galaxy
assembly for some fraction of the high redshift ULIRG population
\citep[e.g.][]{2008ApJ...687...59G}.  Recent theoretical work has suggested
that many SMGs may be produced not by major mergers, but instead by smooth gas
inflow and the accretion of small gas-rich satellites
\citep{2010MNRAS.tmp..360D}.  Such an explanation would be surprising, given
the evidence already in place favoring a major merger origin for SMGs largely
based on dynamical and kinematic arguments \citep[e.g.][]{2005MNRAS.359.1165G,
2006MNRAS.371..465S, 2008ApJ...680..246T,2010ApJ...724..233E}. While there is
no definitive evidence in the data presented here that can unambiguously
support this smooth inflow mode of galaxy formation, the relatively normal
morphologies observed in the non-DOGs could suggest that major mergers are not
responsible for driving the prodigious on-going star-formation in these
systems. Given that such intense star-formation bursts can only be sustained
over a short timescale, the morphologies suggest that the fuel may have to be
accreted in less disruptive minor mergers or through some smooth process.
Physical mechanisms explaining how such a process might occur have been
presented recently \citep{2010ApJ...719..229G,2010arXiv1011.0433G}.
Observations of the internal dynamics of these systems \citep[along the lines
of, e.g.,][]{2008ApJ...687...59G,2010arXiv1011.1507F,2010arXiv1011.5360G} are
likely what is needed to continue progress in this area of research.  
% requisite short timescales over which such bursts can be sustained, yet the
% results presented here provide some indication that this may be happening
% more commonly in some of the bluer (i.e., less dust-obscured) high-redshift
% ULIRGs.

A third possibility is that the expected trends in morphologies with merger
stage are somewhat sensitive both to the initial conditions of the merger ---
for example, highly radial orbits can have similar $G$ and $M_{20}$ values
throughout the ``final merger'' stage \citep{2008MNRAS.391.1137L} --- as well
as the viewing time and angle.  It would be surprising if unusual initial
conditions or viewing times and angles were necessary to explain most high
redshift ULIRGs, particularly since they appear to have fairly typical axial
ratios (see Table~\ref{tab:morph}).  

%A second possibility is that the less obscured sources do not evolve into a
%post-merger state where AGN dominate; perhaps in these objects AGN never even
%form.  This could result from the amount of dust obscuration being related to
%the mass of the host dark matter halo or to the mass of the progenitor merging
%system.

An important consideration related to the XFLS ULIRGs and SMGs analyzed here is
that many of these objects are composite starburst and AGN systems with complex
mid-IR spectral features.  \citet{2008ApJ...680..232D} show that the 7.7$\mu$m
PAH feature is usually strong in extended sources, while it varies from strong
to weak in compact sources.  The mid-IR spectral analysis of these sources
\citep{2007ApJ...664..713S} indicates that only a few XFLS ULIRGs are clearly
dominated by PAH features or AGN continuum emission.  This result is consistent
with the nature of their mid-IR SEDs and underscores the fact that these
objects are composite systems that are not easily classified by either their
mid-IR spectral features or their rest-frame optical morphologies.  Only 7 SMGs
in the sample studied here have both high-resolution imaging and mid-IR
spectroscopy \citep{2009ApJ...699..667M}.  Of these 7, all are bump sources, 4
have strong PAH emission, and 3 have weak or no PAH emission.  It may be the
case that the mid-IR SEDs of the SMG and XFLS ULIRG samples are not
sufficiently distinct to identify significant morphology differences in the
bump vs. power-law sub-samples.

\section{Conclusions} \label{sec:conclusions}

We have used {\it HST} imaging to analyze the morphologies of 22 DOGs at $z
\approx 2$ from the Bo\"{o}tes field selected to show SED features typical of
star-formation dominated systems (bump DOGs).  We compare these new data with
similar {\it HST} imaging of DOGs with SED features typical of AGN-dominated
systems (power-law DOGs), sub-millimeter galaxies (SMGs), and a sample of 
ULIRGs at high-$z$ selected from the {\it Spitzer} XFLS.  Our findings are
summarized below.

\begin{enumerate}
		
\item Spatially resolved emission is observed in the rest-frame optical imaging
    of all bump DOGs.  GALFIT modeling indicates that the point source fraction
    (ratio of flux in the point-source component to total flux of the source)
    in these objects never exceeds 20\% and is typically smaller than that
    found in power-law DOGs, suggesting a smaller AGN contribution to the
    rest-frame optical light from bump DOGs.

\item Typical S\'ersic indices of the resolved emission of bump DOGs suggest
    disk-type rather than bulge-type profiles ($n<2$), similar to power-law
    DOGs.  

\item At $H < 22.5$, bump and power-law DOGs have similar sizes (median $r_{\rm
    P} = 8$~kpc).  At $H > 22.5$, bump DOGs are significantly larger than
    power-law DOGs (median value of $r_{\rm P} =8$~kpc vs. $r_{\rm P} =
    5.4$~kpc, respectively).  This distinction is also true for SMGs and XFLS
    ULIRGs.  

\item In the rest-frame optical, bump DOGs have lower $G$ and higher $M_{20}$
    values than power-law DOGs.  This difference is consistent with
    expectations from simulations of major mergers.  On the other hand, less
    obscured objects in our sample (SMGs and XFLS ULIRGs that do not qualify as
    DOGs) have high $G$ and low $M_{20}$ values that are more typical of
    ``single-object'' systems. 

\end{enumerate}

Overall, our findings highlight the diversity and complexity of high redshift
ULIRG morphologies.  Within the highly obscured subset (i.e., DOGs), we find
evidence in support of a major merger paradigm in which bump DOGs evolve into
power-law DOGs.  Within the less obscured subset (i.e., SMGs and XFLS ULIRGs),
the picture is not as clear.  This may be a result of the timescales over which
obscured and less obscured sources can be observed during a major merger. Alternatively, 
that the intense star-formation in these less-obscured ULIRGs is not the result 
of a recent major merger, and 
may be an indication that more quiescent forms of galaxy assembly are important
for some high redshift ULIRGs.

The work is based primarily on observations made with the {\it Hubble Space
Telescope}.  This work also relies in part on observations made with the {\it
Spitzer Space Telescope}, which is operated by the Jet Propulsion Laboratory,
California Institute of Technology under NASA contract 1407. We are grateful to
the expert assistance of the staff Kitt Peak National Observatory where the
Bo\"{o}tes field observations of the NDWFS were obtained. The authors thank
NOAO for supporting the NOAO Deep Wide-Field Survey. In particular, we thank
Jenna Claver, Lindsey Davis, Alyson Ford, Emma Hogan, Tod Lauer, Lissa Miller,
Erin Ryan, Glenn Tiede and Frank Valdes for their able assistance with the
NDWFS data.  We also thank the staff of the W.~M.~Keck Observatory, where some
of the galaxy redshifts were obtained.

We gratefully acknowledge the anonymous referee whose helpful suggestions have
resulted in an improved manuscript.  RSB gratefully acknowledges financial
assistance from HST grants GO-10890 and GO-11195, without which this research
would not have been possible.  Support for Program numbers HST-GO10890 and
HST-GO11195 were provided by NASA through a grant from the Space Telescope
Science Institute, which  is operated by the Association of Universities for
Research in Astronomy, Incorporated, under NASA contract NAS5-26555.  The
research activities of AD and BTJ are supported by NOAO, which is operated by
the Association of Universities for Research in Astronomy (AURA) under a
cooperative agreement with the National Science Foundation.  Support for E. Le
Floc'h was provided by NASA through the Spitzer Space Telescope Fellowship
Program.

%\clearpage

%\bibliographystyle{apj}

%\bibliography{hstbump}

\begin{appendix}

\section{Images}  \label{sec:galim}

In this section, we present postage stamp images and provide a brief
qualitative description of each of the bump DOGs (as well as one DOG from the
Cycle~16 {\it HST} imaging program that is a power-law source).
Figure~\ref{fig:cutouts1} shows $3\arcsec \times 3\arcsec$ cutout images of the
DOGs in order of increasing redshift (note that redshifts are not available for
the first two sources presented).  Each cutout is centered roughly on the
centroid of emission as seen in the NICMOS image.  A red plus sign shows the
centroid of IRAC 3.6$\mu$m emission and is sized to represent the 1-$\sigma$
uncertainty in the position, which includes independent contributions from the
centroiding error on the 3.6$\mu$m emission ($\approx$0$\farcs$1-0$\farcs$3,
depending on S/N), the relative astrometric calibration uncertainty within the
3.6$\mu$m map ($\approx$0$\farcs$2), and the uncertainty in tying the 3.6$\mu$m
map to the {\it HST} images ($\approx$0$\farcs$1).  The 1$\sigma$ rms offset
between IRAC and NICMOS centroids of the sample is 0$\farcs$2.  In most cases,
the offset in centroids is negligible, but those cases where it is not are
associated with faint 3.6$\mu$m emission (when the absolute astrometric
uncertainty may be as large as 0\farcs4).  This suggests there is no
significant offset between the near-IR and mid-IR centroids at $> 1$~kpc
scales.

The DOGs exhibit a wide range of morphologies, with most being well-resolved.
Only one object (SST24~J143143.3+324944) shows strong Airy rings and is clearly
an unresolved point source.  However, we note that this source has a power-law
dominated mid-IR SED and is not representative of the bump DOG population.
Here we give a brief qualitative description of the morphology of each object.

{\bf (1) SST24 J143143.3+324944:}  F814W: Faint compact morphology.  F160W:
Bright, compact morphology; dominated by unresolved component.

{\bf (2) SST24 J143152.4+350029:}  F814W: Faint diffuse morphology.  F160W:
Bright, extended morphology; low surface brightness extension to southwest.

{\bf (3) SST24 J142724.9+350823:}  F814W: Faint, compact source
$\approx$0$\farcs$5 SW of NIC2 centroid.  F160W: Bright, extended morphology with
tentative evidence of tidal tails or spiral arms.

{\bf (4) SST24 J142951.1+342041:}  F814W: Faint, compact source
$\approx$0$\farcs$3 north of NIC2 centroid.  F160W: Bright, clumpy morphology.

{\bf (5) SST24 J143216.8+335231:}  F814W: Faint, compact source at eastern
edge of NIC2 emission.  F160W: Bright, clumpy morphology; two bad pixels within
the segmentation map of this galaxy have been masked out in the analysis.

{\bf (6) SST24 J143137.1+334500:}  F814W: No detection.  F160W: Extended
narrow morphology resembling a giant edge-on disk with semi-major axis larger
than 3$\arcsec$.

{\bf (7) SST24 J142832.4+340850:}  F814W: Faint, compact morphology.  F160W: No
usable data.

{\bf (8) SST24 J143816.6+333700:}  F814W: Faint, compact morphology.  F160W:
Bright, compact morphology; no obvious PSF signature.

{\bf (9) SST24 J143349.5+334602:}  F814W: Faint, clumpy morphology.  F160W:
Two distinct faint, compact sources; IRAC centroid is closer to eastern source.

{\bf (10) SST24 J143020.4+330344:}  F814W: No detection.  F160W:
Compact morphology; no obvious PSF signature.

{\bf (11) SST24 J142652.4+345504:}  F814W: No detection.  F160W:
Two faint sources separated by $\approx$2$\arcsec$; IRAC centroid consistent
with northeastern source.

{\bf (12) SST24 J142941.0+340915:}  F814W: No detection.  F160W: Clumpy
morphology.

{\bf (13) SST24 J143324.2+334239:}  F814W: Faint, compact morphology.  F160W:
Bright, compact morphology; low surface brightness extension to southwest.

{\bf (14) SST24 J143331.9+352027:}  F814W: Very faint, clumpy morphology.
F160W: Bright, clumpy morphology; low surface brightness extension to
northeast.

{\bf (15) SST24 J143503.3+340243:}  F814W: No detection.  F160W: Bright,
compact morphology; no obvious PSF signature.

{\bf (16) SST24 J142920.1+333023:}  F814W: Faint, compact morphology.  F160W:
Bright, compact morphology.

{\bf (17) SST24 J143321.8+342502:}  F814W: Faint, compact source spatially
coincident with peak NIC2 emission.  F160W:
Bright, compact morphology; no obvious PSF signature; strong low surface
brightness feature extending northeast.

{\bf (18) SST24 J143502.9+342657:}  F814W: No detection.  F160W:
Very clumpy morphology with low surface brightness feature extending to south.

{\bf (19) SST24 J143458.8+333437:}  F814W: Very faint, compact morphology.
F160W: Bright, compact morphology; low surface brightness feature to northwest
resembles a tidal tail.

{\bf (20) SST24 J143028.5+343221:}  F814W: Very faint, clumpy morphology.
F160W: Bright, clumpy morphology; low surface brightness features extending in
eastern and southern directions.

{\bf (21) SST24 J143702.0+344631:}  F814W: No detection.  F160W: No detection.

{\bf (22) SST24 J142637.3+333025:}  F814W: Faint compact morphology.  F160W:
Faint compact morphology; formally detected at 3$\sigma$ level with 0$\farcs$6
diameter aperture.

\begin{figure}[!tbp]
\epsscale{1.0}
\plotone{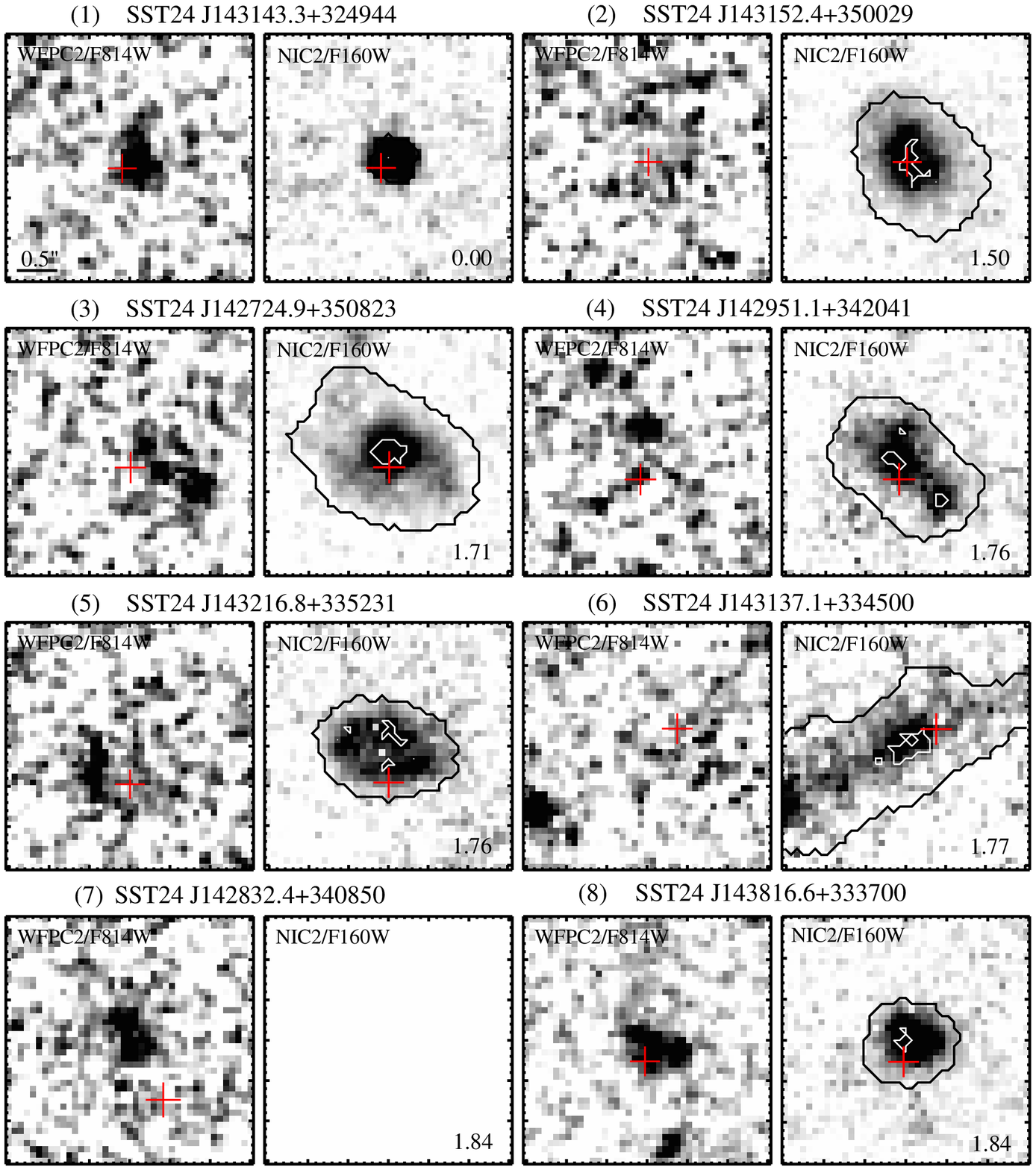}

\caption{\small Cutouts of the 22 DOGs observed by {\it HST}, shown with a
linear stretch.  Columns 1 and 3 are the rest-UV images from WFPC2 F814W and
columns 2 and 4 are the rest-optical images from NIC2 F160W.  Each cutout is
3$\arcsec$ on a side and is oriented north up and east left.  The objects are
arranged in order of increasing redshift, and the redshift is printed in the
lower right corner of each NICMOS image.  A red cross denotes the position and
1-$\sigma$ uncertainty in the centroid of the IRAC 3.6$\mu$m emission.  In
images where the S/N per pixel is greater than 2, white contours outline the
brightest 20\% pixels (for computing $M_{20}$), and black contours show the
outline of the segmentation map used in measuring the non-parametric
morphologies.  NICMOS imaging is not available for target
SST24~J142832.4+340850.}

\label{fig:cutouts1}
\addtocounter{figure}{-1}
\end{figure}

\begin{figure}[!tbp]
\epsscale{1.0}
\plotone{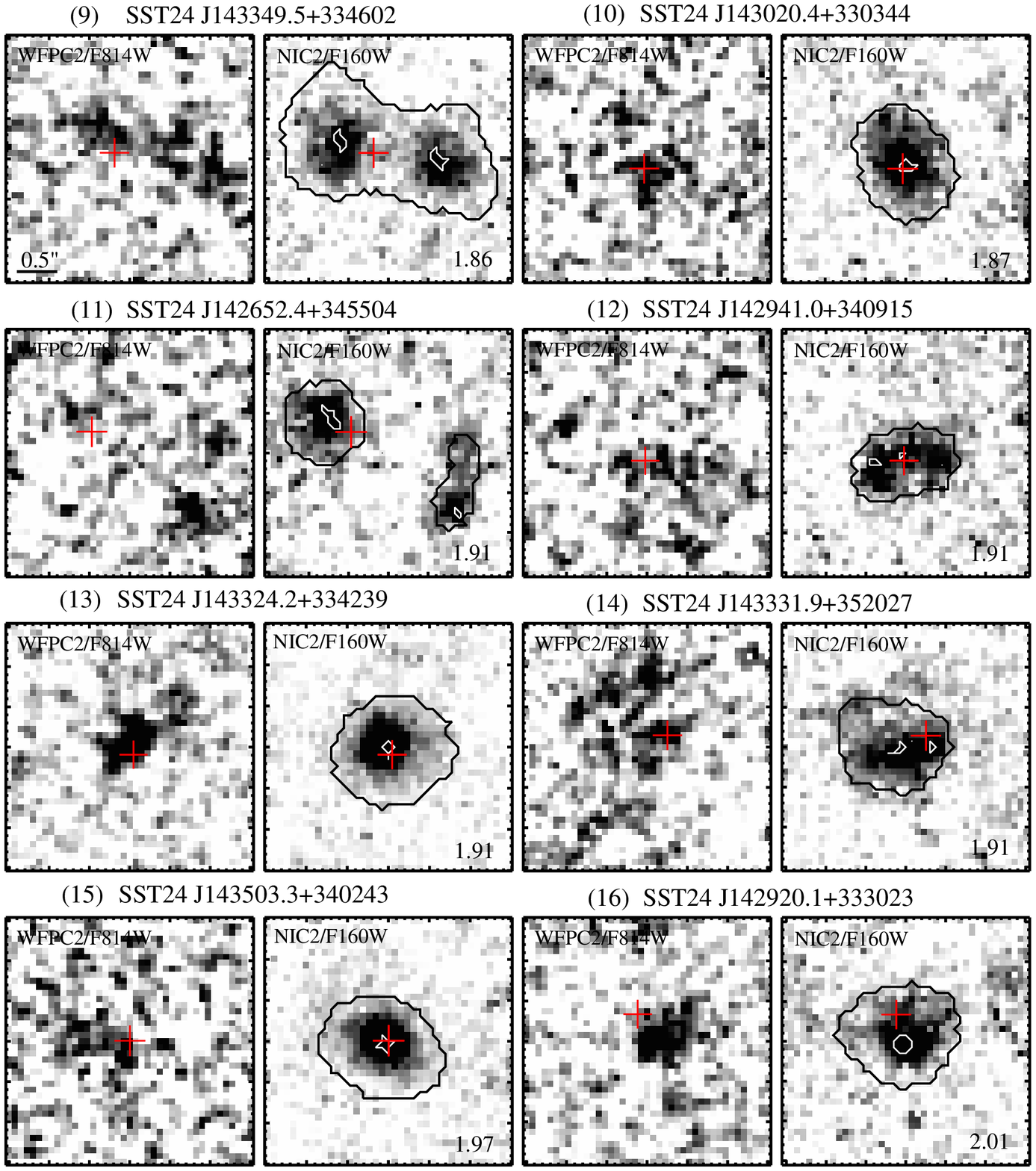}

\caption{Continued.}

\label{fig:cutouts2}
\addtocounter{figure}{-1}
\end{figure}

\begin{figure}[!tbp]
\epsscale{1.0}
\plotone{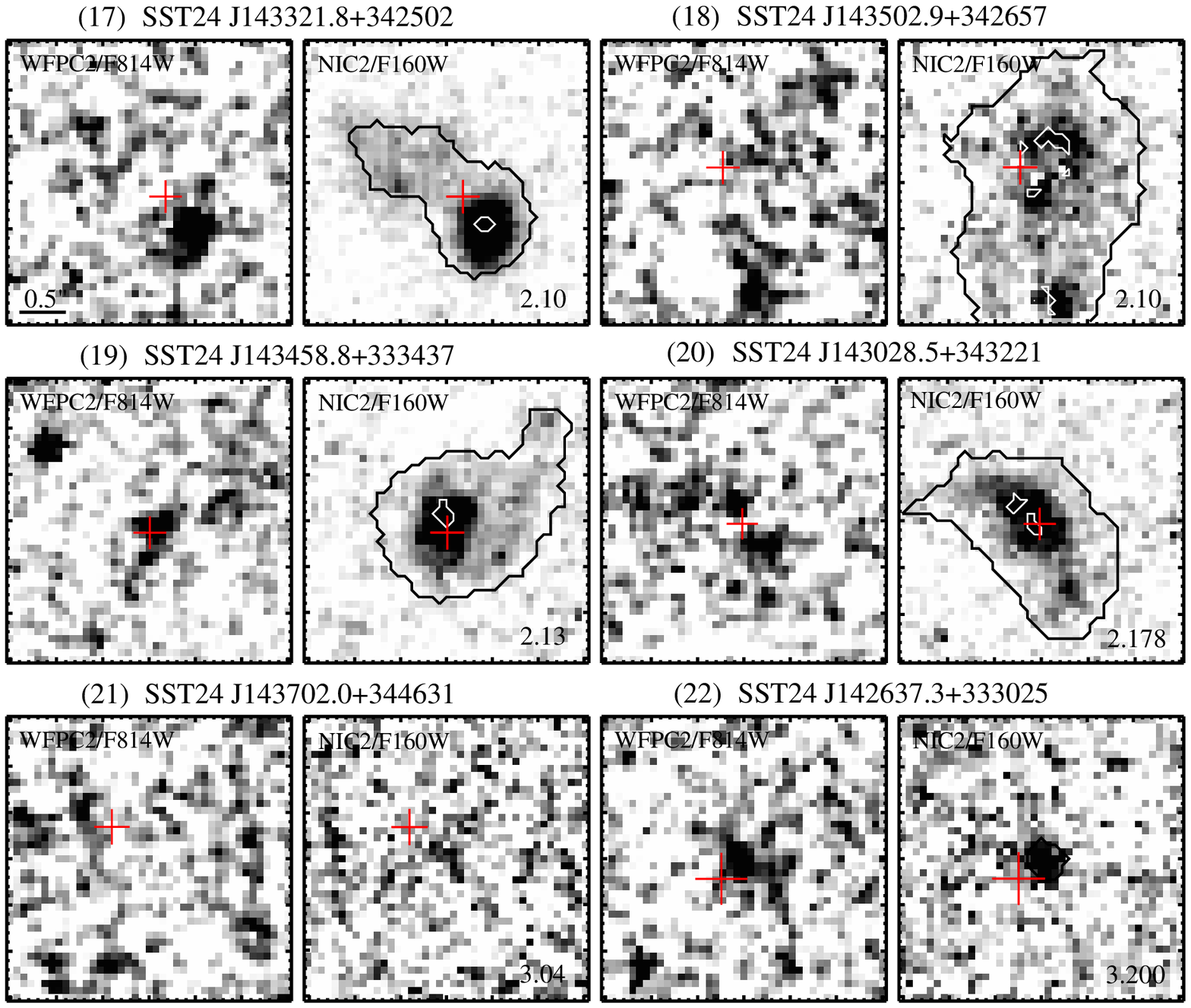}

\caption{Continued.}

\label{fig:cutouts3}
%\addtocounter{figure}{-1}
\end{figure}

\section{SMG and XFLS ULIRG Non-parametric Morphologies}\label{sec:smgmorph}

The morphologies presented herein comprise a large sample of high redshift
ULIRGs analyzed in a uniform manner.  This minimizes systematic uncertainties
in the morphological measurements by facilitating interpretation of the results
in a relative sense.

Table~\ref{tab:smgmorph} presents the measurements of non-parametric
morphologies of SMGs at $z > 1.4$ derived from NIC2 images using the same
morphology code used to analyze the imaging of XFLS ULIRGs and DOGs.  A total
of 18 SMGs meet this requirement, but 2 of these have per-pixel-S/N$<2$ and are
not included in our analysis here.  This table also includes an estimate of
whether the source is dominated by a bump or by a power-law in the mid-IR using
IRAC data from \citet{2009ApJ...699.1610H} and the same statistical definition
originally used for DOGs \citep{2008ApJ...677..943D}.

\citet{2010MNRAS.405..234S} present measurements of $r_{\rm P}$ and $G$ for
SMGs, and it is instructive to compare their results with ours here.  We find
that our size measurements are generally consistent, with median $r_{\rm P}$
values of 8.4~kpc in our analysis and 8.6~kpc in that of
\citet{2010MNRAS.405..234S}.  We also find no systematic offset either at large
or small radii in the $r_{\rm P}$ values.

On the other hand, we find significant offsets in the respective measurements
of $G$.  Our median $G$ value for SMGs at $z>1.4$ is 0.49, while that of
\citet{2010MNRAS.405..234S} is 0.54.  Additionally, aside from a few
exceptions, there is tentative evidence that the offset increases with
S/N-per-pixel.  These offsets may be the result of a different means of
selecting which pixels belong to the galaxy in question.  As discussed in
section~\ref{sec:nonparmeth}, pixels with surface brightness above $\mu(r_{\rm
P})$ are assigned to the galaxy while those below it are not.  Meanwhile,
\citet{2010MNRAS.405..234S} adopt 1.5$r_{\rm P}$ as their Petrosian radius.
Studies of the morphologies of galaxies in the {\it HST} Ultra-deep field (UDF)
have shown that the $G$ coefficient has a strong dependence on the specific
definition used for the Petrosian radius \citep{2008ApJS..179..319L}.  At
reliable S/N levels (S/N$>2$), \citet{2008ApJS..179..319L} show that using the
larger aperture to define a galaxy's extent can cause an increase in $G$ of up
to 0.1, with some evidence for an increase in the offset with S/N.  This effect
is thus qualitatively consistent with the differences observed between our
measurements and those presented in \citet{2010MNRAS.405..234S}.

The primary takeaway of this comparison is that when comparing morphologies of
objects, it is necessary to apply a single systematic method in analyzing all
objects in the sample.  We note that the central conclusions presented in
\citet{2010MNRAS.405..234S} are based on measurements of the morphologies of
SMGs relative to a population of field galaxies and are therefore robust.

Finally, Table~\ref{tab:xflsmorph} presents our measurements of non-parametric
morphologies of XFLS ULIRGs at $z > 1.4$ derived from NIC2 images using the
same morphology code used to analyze the imaging of SMGs and DOGs.  

\clearpage

\begin{landscape}

\begin{deluxetable*}{llccccccccccccc}

%\rotate
\tabletypesize{\tiny} 
\tablecolumns{15}
\tablewidth{8.5in}
\tablecaption{NICMOS Morphological Classifications}
\tablehead{
\colhead{} & \colhead{} & \colhead{} & \colhead{} & \colhead{} & \colhead{$r_{\rm P}$} & \colhead{} & \colhead{} & \colhead{} & 
\colhead{} & \colhead{$R_{\rm eff}$} & \colhead{} & \colhead{} & \colhead{} & \colhead{} \\
\colhead{Source Name} & 
\colhead{SED} & 
\colhead{Visual\tablenotemark{a}} & 
\colhead{$N_{\rm agr}$\tablenotemark{b}} & 
\colhead{$S/N$} &
\colhead{(kpc)} & 
\colhead{$G$} & 
\colhead{$M_{20}$} & 
\colhead{$C$} & 
\colhead{PSF Fraction} & \colhead{(kpc)} & \colhead{$n$} & \colhead{Axial Ratio} & 
\colhead{$N_{\rm dof}$} & \colhead{$\chi^2_\nu$}
}
\startdata
SST24 J142637.3+333025 & PL   & TFTT& 4 &  3.1 &  2.3$\pm$1.0 & 0.47$\pm$0.03& -1.72$\pm$0.10 & 2.9$\pm$0.4&  0.45$\pm$0.55 & ---         &   0.1$\pm$1.7  &  0.15$\pm$0.49  & 1594  & 2.3   \\
SST24 J142652.4+345504 & Bump & Reg & 4 &  3.8 &  8.6$\pm$0.9 & 0.38$\pm$0.03& -0.77$\pm$0.10 & 3.4$\pm$0.4&  0.11$\pm$0.20 & 3.3$\pm$0.4 &   1.1$\pm$0.2  &  0.84$\pm$0.06  & 1654  & 1.6   \\
SST24 J142724.9+350823 & Bump & Irr & 6 &  4.1 & 12.0$\pm$0.8 & 0.48$\pm$0.03& -1.63$\pm$0.10 & 4.9$\pm$0.4&  0.04$\pm$0.06 & 4.6$\pm$0.2 &   1.5$\pm$0.1  &  0.69$\pm$0.02  & 1656  & 1.1   \\
SST24 J142832.4+340850 & Bump & --- &---&  --- &  ---         &      ---     &    ---         &    ---     &            --- &  --         &  ---           & ---             & ---   & ---   \\
SST24 J142920.1+333023 & Bump & Irr & 6 &  3.4 &  6.9$\pm$1.0 & 0.48$\pm$0.03& -1.00$\pm$0.10 & 2.8$\pm$0.4&  0.08$\pm$0.09 & 3.3$\pm$0.2 &   0.8$\pm$0.1  &  0.79$\pm$0.04  & 1663  & 2.5   \\
SST24 J142941.0+340915 & Bump & Irr & 7 &  2.8 &  6.6$\pm$1.2 & 0.40$\pm$0.04& -0.99$\pm$0.11 & 1.7$\pm$0.5&  0.07$\pm$0.18 & 3.9$\pm$1.8 &   0.0$\pm$0.1  &  0.57$\pm$0.04  & 1635  & 2.1   \\
SST24 J142951.1+342041 & Bump & Irr & 6 &  4.0 & 10.7$\pm$0.8 & 0.46$\pm$0.03& -0.98$\pm$0.10 & 2.8$\pm$0.4&  0.03$\pm$0.09 & 5.2$\pm$0.1 &   0.3$\pm$0.1  &  0.42$\pm$0.01  & 1668  & 3.2   \\
SST24 J143020.4+330344 & Bump & Reg & 7 &  3.3 &  7.0$\pm$1.0 & 0.49$\pm$0.03& -1.63$\pm$0.10 & 3.0$\pm$0.4&  0.13$\pm$0.11 & 2.9$\pm$0.2 &   0.9$\pm$0.2  &  0.64$\pm$0.04  & 1599  & 3.4   \\
SST24 J143028.5+343221 & PL   & Irr & 6 &  2.5 &  9.7$\pm$1.3 & 0.51$\pm$0.05& -1.18$\pm$0.13 & 4.0$\pm$0.5&  0.05$\pm$0.05 & 4.4$\pm$0.2 &   0.7$\pm$0.1  &  0.39$\pm$0.02  & 1639  & 2.8   \\
SST24 J143137.1+334500 & Bump & Irr & 4 &  2.5 & 27.5$\pm$1.3 & 0.44$\pm$0.05& -1.00$\pm$0.13 & 3.2$\pm$0.5&  0.10$\pm$0.13 &10.5$\pm$0.8 &   0.7$\pm$0.1  &  0.26$\pm$0.01  & 1657  & 1.1   \\
SST24 J143143.3+324944 & PL   & Reg & 7 & 11.3 &  0.0$\pm$0.0 & 0.52$\pm$0.02& -1.69$\pm$0.06 & 3.0$\pm$0.3&  0.42$\pm$0.05 & ---         &   1.0$\pm$0.4  &  0.66$\pm$0.03  & 1665  & 2.7   \\
SST24 J143152.4+350029 & Bump & Irr & 4 &  5.2 &  8.9$\pm$0.8 & 0.46$\pm$0.03& -1.41$\pm$0.06 & 4.8$\pm$0.4&  0.02$\pm$0.03 & 3.9$\pm$0.1 &   0.7$\pm$0.1  &  0.69$\pm$0.01  & 1666  & 2.2   \\
SST24 J143216.8+335231 & Bump & Irr & 4 &  4.4 &  8.4$\pm$0.8 & 0.38$\pm$0.03& -0.98$\pm$0.08 & 2.4$\pm$0.4&  0.00$\pm$0.08 & 4.2$\pm$0.1 &   0.1$\pm$0.1  &  0.60$\pm$0.01  & 1666  & 2.0   \\
SST24 J143321.8+342502 & Bump & Irr & 6 &  5.0 &  8.2$\pm$0.8 & 0.54$\pm$0.03& -0.78$\pm$0.06 & 3.3$\pm$0.4&  0.13$\pm$0.05 & 1.8$\pm$0.1 &   1.4$\pm$0.1  &  0.61$\pm$0.02  & 1659  & 1.6   \\
SST24 J143324.2+334239 & Bump & Reg & 7 &  3.8 &  6.7$\pm$0.9 & 0.54$\pm$0.03& -1.62$\pm$0.10 & 3.0$\pm$0.4&  0.12$\pm$0.06 & 2.6$\pm$0.1 &   1.2$\pm$0.1  &  0.80$\pm$0.03  & 1612  & 2.6   \\
SST24 J143331.9+352027 & Bump & Irr & 6 &  4.1 &  8.2$\pm$0.8 & 0.37$\pm$0.03& -0.85$\pm$0.09 & 2.4$\pm$0.4&  0.00$\pm$0.08 & 4.7$\pm$0.2 &   0.9$\pm$0.1  &  0.57$\pm$0.03  & 1658  & 2.7   \\
SST24 J143349.5+334602 & Bump & Irr & 7 &  2.5 & 11.8$\pm$1.3 & 0.48$\pm$0.05& -0.83$\pm$0.12 & 1.9$\pm$0.5&  0.05$\pm$0.07 & 3.2$\pm$0.3 &   0.8$\pm$0.1  &  0.67$\pm$0.04  & 1660  & 2.4   \\
SST24 J143458.8+333437 & Bump & Irr & 7 &  2.5 &  9.6$\pm$1.3 & 0.54$\pm$0.05& -1.24$\pm$0.12 & 3.8$\pm$0.5&  0.15$\pm$0.10 & 4.9$\pm$0.5 &   2.1$\pm$0.3  &  0.81$\pm$0.03  & 1657  & 1.9   \\
SST24 J143502.9+342657 & Bump & Irr & 7 &  2.1 & 14.9$\pm$1.5 & 0.46$\pm$0.05& -0.77$\pm$0.15 & 2.4$\pm$0.6&  0.03$\pm$0.25 & 8.8$\pm$3.5 &   0.0$\pm$0.1  &  0.35$\pm$0.02  & 1669  & 1.3   \\
SST24 J143503.3+340243 & Bump & Reg & 6 &  4.0 &  7.2$\pm$0.8 & 0.53$\pm$0.03& -1.71$\pm$0.10 & 2.8$\pm$0.4&  0.09$\pm$0.09 & 2.9$\pm$0.1 &   1.0$\pm$0.1  &  0.62$\pm$0.02  & 1659  & 1.8   \\
SST24 J143702.0+344631 & Bump & TFTT& 7 &  --- &     ---      &    ---       &     ---        &    ---     &       ---      &     ---     &        ---     &     --          &  ---  & ---   \\
SST24 J143816.6+333700 & Bump & Reg & 5 &  5.1 &  5.7$\pm$0.8 & 0.47$\pm$0.03& -1.47$\pm$0.06 & 2.6$\pm$0.4&  0.06$\pm$0.04 & 2.3$\pm$0.1 &   0.4$\pm$0.1  &  0.81$\pm$0.02  & 1666  & 1.9   \\
\enddata
\tablenotetext{a}{ Mode of visual classification.}
\tablenotetext{b}{ Number of users in agreement with mode of visual classification.}
\label{tab:morph}
%\end{landscape}
\end{deluxetable*}

\clearpage

\begin{deluxetable*}{llcccccccccccc}
    %\begin{landscape}
%\rotate
\tabletypesize{\scriptsize} 
\tablecolumns{14}
\tablewidth{8.5in}
\tablecaption{SMG NICMOS Morphological Classifications}
\tablehead{
\colhead{} &\colhead{} &\colhead{} &\colhead{$r_{\rm P}$} &\colhead{} &
\colhead{} &\colhead{} & \colhead{} & \colhead{$R_{\rm eff}$} & \colhead{} &  
\colhead{} & \colhead{} & \colhead{} \\
\colhead{Source Name} & 
\colhead{SED} & 
\colhead{$S/N$} &
\colhead{(kpc)} & 
\colhead{$G$} & 
\colhead{$M_{20}$} & 
\colhead{$C$} &
\colhead{PSF Fraction} & \colhead{(kpc)} & \colhead{$n$} &  
\colhead{$N_{\rm dof}$} & \colhead{$\chi^2_\nu$}
}
\startdata
  CFRS03-15 &  Bump &  6.1 & 12.5$\pm$0.8 &  0.57$\pm$0.02 & -1.72$\pm$0.06 &  4.3$\pm$0.3 &  0.00$\pm$ 0.02 &  40.7$\pm$34.1 &  18.3$\pm$3.8  & 1671 &    4.3 \\
 LOCKMAN-03 &  Bump &  4.4 & 13.4$\pm$0.8 &  0.51$\pm$0.03 & -1.15$\pm$0.08 &  3.0$\pm$0.4 &  0.00$\pm$ 0.03 &   4.5$\pm$ 0.2 &   1.2$\pm$0.1  & 1671 &    1.4 \\
 LOCKMAN-06 &  Bump &  3.5 & 10.2$\pm$0.9 &  0.48$\pm$0.03 & -1.46$\pm$0.10 &  3.0$\pm$0.4 &  0.00$\pm$ 0.05 &   5.3$\pm$ 0.4 &   2.0$\pm$0.1  & 1671 &    0.9 \\
 LOCKMAN-02 &  Bump &  4.0 & 12.9$\pm$0.8 &  0.46$\pm$0.03 & -0.99$\pm$0.10 &  4.5$\pm$0.4 &  0.03$\pm$ 0.07 &   5.4$\pm$ 0.1 &   0.5$\pm$0.0  & 1663 &    0.7 \\
   HDFN-082 &  Bump & $<2$ &     ---      &      ---       &      ---       &     ---      &       ---       &       ---      &       ---      &  --- &    --- \\
   HDFN-092 &  Bump &  2.0 &  8.2$\pm$1.4 &  0.45$\pm$0.05 & -0.96$\pm$0.15 &  5.4$\pm$0.6 &  0.11$\pm$ 0.16 &   4.1$\pm$ 0.1 &   0.1$\pm$0.1  & 1671 &    1.2 \\
   HDFN-093 &  Bump &  5.8 &  3.4$\pm$0.8 &  0.49$\pm$0.02 & -1.76$\pm$0.06 &  3.2$\pm$0.3 &  0.19$\pm$ 0.03 &   1.3$\pm$ 0.2 &   4.2$\pm$1.2  & 1664 &    0.7 \\  
   HDFN-105 &  Bump &  7.1 &  4.8$\pm$0.7 &  0.49$\pm$0.02 & -1.73$\pm$0.06 &  2.8$\pm$0.3 &  0.00$\pm$ 0.08 &   2.5$\pm$ 0.2 &   2.9$\pm$0.2  & 1671 &    1.4 \\
   HDFN-127 &  PL   &  3.1 &  4.6$\pm$1.0 &  0.49$\pm$0.03 & -1.17$\pm$0.10 &  3.5$\pm$0.4 &  0.41$\pm$ 0.21 &   1.7$\pm$ 0.2 &   0.2$\pm$0.2  & 1671 &    1.4 \\
   HDFN-143 &  Bump &  3.4 &  8.4$\pm$1.0 &  0.34$\pm$0.03 & -1.04$\pm$0.10 &  2.4$\pm$0.4 &  0.02$\pm$ 0.08 &   4.4$\pm$ 0.1 &   0.2$\pm$0.1  & 1670 &    0.7 \\
   HDFN-161 &  Bump &  5.5 &  5.4$\pm$0.8 &  0.58$\pm$0.03 & -1.80$\pm$0.06 &  3.4$\pm$0.3 &  0.01$\pm$ 0.10 &  29.8$\pm$41.9 &  20.0$\pm$7.6  & 1671 &    1.9 \\
   HDFN-172 &  Bump &  5.5 &  8.6$\pm$0.8 &  0.46$\pm$0.03 & -1.02$\pm$0.06 &  2.4$\pm$0.3 &  0.18$\pm$ 0.09 &   4.1$\pm$ 0.3 &   1.7$\pm$0.2  & 1671 &    1.7 \\
   SA13-332 &  PL   &  5.1 &  3.2$\pm$0.7 &  0.51$\pm$0.03 & -1.62$\pm$0.06 &  3.0$\pm$0.4 &  0.47$\pm$ 0.03 &   1.4$\pm$ 0.1 &   0.9$\pm$0.3  & 1666 &    0.6 \\
   SA13-570 &  PL   &  3.2 &  7.0$\pm$1.0 &  0.49$\pm$0.03 & -1.76$\pm$0.10 &  2.7$\pm$0.4 &  0.03$\pm$ 0.18 &   2.9$\pm$ 0.1 &   1.5$\pm$0.2  & 1662 &    0.5 \\
   CFRS14-3 &  Bump &  5.7 &  6.1$\pm$0.8 &  0.59$\pm$0.02 & -1.56$\pm$0.06 &  3.4$\pm$0.3 &  0.01$\pm$ 0.09 &   1.6$\pm$ 0.1 &   3.4$\pm$0.3  & 1671 &    1.4 \\
   ELAIS-13 &  Bump & $<2$ &      ---     &          ---   &      ---       &     ---      &       ---       &       ---      &       ---      &  --- &    --- \\
   ELAIS-07 &  Bump &  4.7 &  8.6$\pm$0.8 &  0.46$\pm$0.03 & -0.96$\pm$0.07 &  4.3$\pm$0.4 &  0.00$\pm$ 0.08 &   3.7$\pm$ 0.2 &   1.4$\pm$0.1  & 1671 &    1.5 \\
   ELAIS-04 &  Bump &  5.6 &  9.3$\pm$0.8 &  0.54$\pm$0.03 & -1.30$\pm$0.06 &  3.8$\pm$0.3 &  0.05$\pm$ 0.04 &   3.3$\pm$ 0.0 &   0.5$\pm$0.0  & 1671 &    3.3 \\
\enddata
\label{tab:smgmorph}
%\end{landscape}
\end{deluxetable*}

\clearpage

\begin{deluxetable*}{llcccccccccccc}
    %\begin{landscape}
%\rotate
\tabletypesize{\tiny} 
\tablecolumns{14}
\tablewidth{6.5in}
\tablecaption{XFLS NICMOS Morphological Classifications}
\tablehead{
\colhead{} &\colhead{} &\colhead{} &\colhead{$r_{\rm P}$} &\colhead{} &
\colhead{} &\colhead{} & \colhead{} & \colhead{$R_{\rm eff}$} & \colhead{} &  
\colhead{} & \colhead{} & \colhead{} \\
\colhead{Source Name} & 
\colhead{SED} & 
\colhead{$S/N$} &
\colhead{(kpc)} & 
\colhead{$G$} & 
\colhead{$M_{20}$} & 
\colhead{$C$} &
\colhead{PSF Fraction} & \colhead{(kpc)} & \colhead{$n$} &  
\colhead{$N_{\rm dof}$} & \colhead{$\chi^2_\nu$}
}
\startdata
   MIPS506   & Bump &  5.0 &   5.2$\pm$0.8 &  0.46$\pm$0.02 & -1.45$\pm$0.06 &  3.1$\pm$0.3 &   0.15$\pm$0.24   & 2.6$\pm$0.2 &   1.3$\pm$0.1 & 1671  &  0.8 \\
   MIPS289   & Bump &  5.2 &  11.1$\pm$0.8 &  0.54$\pm$0.02 & -2.01$\pm$0.06 &  3.4$\pm$0.3 &   0.08$\pm$0.03   & 4.7$\pm$0.3 &   2.2$\pm$0.2 & 1671  &  2.1 \\
  MIPS8342   & Bump &  8.0 &   5.3$\pm$0.8 &  0.57$\pm$0.03 & -1.73$\pm$0.10 &  3.1$\pm$0.4 &   0.11$\pm$0.04   & 1.3$\pm$0.1 &   2.1$\pm$0.1 & 1671  &  1.2 \\
  MIPS8242   & Bump &  4.7 &  12.8$\pm$0.9 &  0.44$\pm$0.03 & -0.89$\pm$0.10 &  3.3$\pm$0.4 &   0.05$\pm$0.04   & 5.4$\pm$0.1 &   0.5$\pm$0.1 & 1671  &  2.1 \\
   MIPS464   & PL   &  5.2 &   4.6$\pm$0.8 &  0.40$\pm$0.03 & -1.59$\pm$0.06 &  2.5$\pm$0.4 &   0.16$\pm$0.70   & 1.9$\pm$0.1 &   0.7$\pm$0.1 & 1671  &  1.0 \\
   MIPS227   & Bump & 10.4 &   7.7$\pm$1.3 &  0.54$\pm$0.05 & -1.84$\pm$0.13 &  3.0$\pm$0.5 &   0.03$\pm$0.01   & 2.6$\pm$0.1 &   1.6$\pm$0.1 & 1671  &  1.7 \\
  MIPS8196   & Bump &  8.5 &   9.0$\pm$0.7 &  0.54$\pm$0.02 & -2.09$\pm$0.06 &  3.7$\pm$0.3 &   0.07$\pm$0.01   & 4.4$\pm$0.2 &   4.0$\pm$0.2 & 1671  &  1.8 \\
  MIPS8327   & Bump &  5.9 &   5.6$\pm$0.9 &  0.51$\pm$0.03 & -1.44$\pm$0.10 &  2.8$\pm$0.4 &   0.00$\pm$0.06   & 1.8$\pm$0.1 &   3.4$\pm$0.4 & 1671  &  1.4 \\
  MIPS8245   & Bump &  3.2 &   3.5$\pm$0.8 &  0.44$\pm$0.03 & -0.96$\pm$0.06 &  2.2$\pm$0.4 &   0.00$\pm$1.00   & 1.6$\pm$0.1 &   0.4$\pm$0.2 & 1670  &  1.7 \\
    MIPS78   & PL   &  2.2 &   6.5$\pm$0.8 &  0.43$\pm$0.03 & -0.84$\pm$0.06 &  2.5$\pm$0.4 &   0.21$\pm$0.51   & 2.7$\pm$0.3 &   0.3$\pm$0.1 & 1671  &  1.5 \\
   MIPS180   & Bump &  4.7 &   3.6$\pm$0.8 &  0.41$\pm$0.02 & -1.90$\pm$0.06 &  2.4$\pm$0.3 &   0.31$\pm$0.82   & 1.6$\pm$0.1 &   0.2$\pm$0.2 & 1671  &  1.8 \\
    MIPS42   & PL   &  3.4 &   5.2$\pm$0.8 &  0.47$\pm$0.02 & -0.95$\pm$0.06 &  2.5$\pm$0.3 &   0.14$\pm$0.18   & 2.2$\pm$0.3 &   1.1$\pm$0.5 & 1671  &  2.0 \\
  MIPS8493   & Bump &  3.7 &  12.1$\pm$1.5 &  0.49$\pm$0.05 & -1.09$\pm$0.15 &  3.7$\pm$0.6 &   0.00$\pm$0.07   & 5.3$\pm$0.3 &   1.2$\pm$0.1 & 1671  &  1.3 \\
 MIPS22661   & Bump &  8.1 &   4.8$\pm$0.8 &  0.50$\pm$0.03 & -1.81$\pm$0.06 &  2.9$\pm$0.4 &   0.21$\pm$0.04   & 1.8$\pm$0.1 &   1.0$\pm$0.1 & 1670  &  2.4 \\
 MIPS22277   & Bump &  7.8 &   5.9$\pm$0.8 &  0.53$\pm$0.02 & -1.67$\pm$0.06 &  3.0$\pm$0.3 &   0.06$\pm$0.03   & 2.0$\pm$0.1 &   2.0$\pm$0.1 & 1670  &  1.4 \\
 MIPS22204   & PL   & 11.6 &   3.4$\pm$1.0 &  0.51$\pm$0.03 & -1.60$\pm$0.10 &  2.9$\pm$0.4 &   0.17$\pm$0.04   & 0.7$\pm$0.1 &   3.7$\pm$0.3 & 1671  &  1.5 \\
 MIPS16080   & Bump &  5.5 &   9.4$\pm$0.8 &  0.57$\pm$0.03 & -1.39$\pm$0.06 &  3.5$\pm$0.3 &   0.03$\pm$0.03   & 2.8$\pm$0.1 &   2.8$\pm$0.2 & 1671  &  1.4 \\
 MIPS22303   & PL   &  2.4 &   6.4$\pm$0.8 &  0.42$\pm$0.02 & -0.99$\pm$0.06 &  2.6$\pm$0.3 &   0.19$\pm$0.29   & 2.6$\pm$0.5 &   1.1$\pm$0.3 & 1669  &  1.0 \\
 MIPS15977   & Bump &  8.6 &   5.8$\pm$1.4 &  0.52$\pm$0.05 & -1.87$\pm$0.13 &  3.0$\pm$0.5 &   0.22$\pm$0.04   & 2.5$\pm$0.1 &   0.7$\pm$0.1 & 1669  &  1.4 \\
 MIPS15928   & Bump &  7.7 &   7.5$\pm$0.7 &  0.52$\pm$0.02 & -1.90$\pm$0.06 &  3.1$\pm$0.3 &   0.22$\pm$0.05   & 3.2$\pm$0.1 &   0.6$\pm$0.1 & 1671  &  2.8 \\
 MIPS15840   & PL   &  4.4 &   4.8$\pm$0.8 &  0.45$\pm$0.03 & -1.47$\pm$0.10 &  2.8$\pm$0.4 &   0.18$\pm$0.22   & 2.2$\pm$0.1 &   0.6$\pm$0.1 & 1671  &  1.1 \\
 MIPS22651   & Bump &  6.0 &   7.7$\pm$1.4 &  0.58$\pm$0.05 & -2.00$\pm$0.14 &  3.3$\pm$0.6 &   0.11$\pm$0.06   & 2.4$\pm$0.1 &   1.6$\pm$0.1 & 1671  &  1.2 \\
 MIPS22558   & Bump &  4.8 &   3.6$\pm$0.8 &  0.51$\pm$0.03 & -1.84$\pm$0.07 &  3.4$\pm$0.4 &   0.16$\pm$0.12   & 3.1$\pm$2.2 &  10.9$\pm$5.2 & 1671  &  0.8 \\
 MIPS22699   & PL   &  4.3 &   3.6$\pm$1.3 &  0.49$\pm$0.05 & -2.37$\pm$0.13 &  3.0$\pm$0.5 &   0.09$\pm$1.00   & 0.9$\pm$0.1 &   3.2$\pm$0.9 & 1671  &  1.1 \\
 MIPS16122   & PL   &  2.4 &   7.6$\pm$0.7 &  0.46$\pm$0.02 & -1.26$\pm$0.06 &  3.0$\pm$0.3 &   0.04$\pm$0.20   & 2.3$\pm$0.1 &   1.2$\pm$0.1 & 1671  &  1.6 \\
 MIPS15949   & Bump &  4.0 &   8.6$\pm$0.7 &  0.61$\pm$0.02 & -1.52$\pm$0.06 &  3.7$\pm$0.3 &   0.28$\pm$0.05   & 2.8$\pm$0.2 &   1.7$\pm$0.2 & 1671  &  1.0 \\
 MIPS15880   & Bump &  4.0 &   8.8$\pm$1.0 &  0.46$\pm$0.03 & -1.08$\pm$0.10 &  2.3$\pm$0.4 &   0.03$\pm$0.08   & 5.4$\pm$0.3 &   0.7$\pm$0.1 & 1671  &  1.8 \\
 MIPS16113   & Bump &  1.6 &   9.0$\pm$0.8 &  0.47$\pm$0.03 & -0.66$\pm$0.09 &  1.8$\pm$0.4 &   0.02$\pm$0.12   & 2.6$\pm$0.2 &   1.5$\pm$0.2 & 1671  &  1.6 \\
 MIPS22530   & Bump &  2.4 &  10.0$\pm$0.8 &  0.47$\pm$0.03 & -1.42$\pm$0.07 &  2.5$\pm$0.4 &   0.03$\pm$0.13   & 3.9$\pm$0.3 &   0.9$\pm$0.1 & 1664  &  1.5 \\
 MIPS15958   & PL   &  7.7 &   3.7$\pm$0.8 &  0.53$\pm$0.03 & -1.74$\pm$0.08 &  3.0$\pm$0.4 &   0.67$\pm$0.10   & 1.6$\pm$0.1 &   0.4$\pm$0.1 & 1671  &  1.1 \\
 MIPS16095   & Bump &  9.3 &   5.7$\pm$0.8 &  0.52$\pm$0.02 & -1.83$\pm$0.06 &  3.1$\pm$0.3 &   0.06$\pm$0.04   & 1.9$\pm$0.1 &   1.7$\pm$0.1 & 1671  &  1.2 \\
 MIPS16144   & Bump &  3.7 &  13.0$\pm$0.8 &  0.50$\pm$0.02 & -1.46$\pm$0.06 &  4.4$\pm$0.3 &   0.10$\pm$0.04   & 5.6$\pm$1.0 &   3.1$\pm$0.6 & 1664  &  1.3 \\
 MIPS16059   & Bump &  5.2 &   8.4$\pm$0.8 &  0.53$\pm$0.02 & -1.31$\pm$0.06 &  2.5$\pm$0.3 &   0.05$\pm$0.05   & 2.9$\pm$0.1 &   0.3$\pm$0.1 & 1671  &  1.7 \\
\enddata
\label{tab:xflsmorph}
%\end{landscape}
\end{deluxetable*}

\clearpage

\end{landscape}

%\end{sidewaystable}

\end{appendix}

\end{document}